\documentclass[english,preprint,prd]{revtex4}
\usepackage[T1]{fontenc}
\usepackage[latin1]{inputenc}
\usepackage{amsmath}
\usepackage{graphicx}
\usepackage{amssymb}
\makeatletter
\providecommand{\tabularnewline}{\\}
\usepackage{graphicx}
\usepackage{dcolumn}
\usepackage{bm}
\usepackage{amsfonts}
\usepackage{babel}
\makeatother
\providecommand{\stau}{\tilde{\tau}}

\begin{document}
\preprint{CERN-PH-TH/2007-051}
\title{Study of Metastable Staus at Linear Colliders}
\author{Orhan \c{C}ak{\i}r$^1$, \.Ilkay T. \c{C}ak{\i}r$^1$, John R. Ellis$^2$, Zerrin K{\i}rca$^3$}
\affiliation{$^1$Ankara University, Faculty of Sciences, Department of Physics,
06100, Tandogan, Ankara, Turkey.\\
$^2$Theory Division, Physics Department, CERN, CH-1211 Geneva 23,
Switzerland.\\
$^3$Uludag University, Faculty of Arts and Sciences, Department
of Physics, Bursa, Turkey.}

\begin{abstract}
We consider scenarios in which the lightest sparticle (LSP) is
the gravitino and the next-to-lightest sparticle (NLSP) is a metastable stau.
We examine the production of stau pairs in $e^{+}e^{-}$ annihilation
at ILC and CLIC energies. In addition to three minimal supergravity (mSUGRA)
benchmark scenarios proposed previously, we consider a new high-mass
scenario in which effects catalyzed by stau bound states yield abundances of
$^{6,7}$Li that fit the astrophysical data better than standard Big-Bang
nucleosynthesis. This scenario could be probed only at CLIC energies.
In each scenario, we show how the stau mixing angle may be determined
from measurements of the total stau pair-production cross sections with polarized
beams, and of the tau polarization in stau decays.
Using realistic ILC and CLIC luminosity spectra, we find for each scenario
the center-of-mass energy that maximizes the number of staus with $\beta \gamma < 0.4$,
that may be trapped in a generic detector. The dominant sources of such slow-moving
staus are generically the pair production and cascade decays of heavier sparticles
with higher thresholds, and the optimal center-of-mass energy is typically considerably
beyond $2 m_{\tilde\tau_1}$.
\\
\bigskip
\\
\\
March 2008
\end{abstract}

\maketitle

\section{Introduction}

Supersymmetry (SUSY) is one of the most interesting possibilities offered
by quantum field theory whose relevance to nature has not yet been established~\cite{Yao06}.
In addition to relating fermions to bosons, stabilizing the mass scale of the
weak interactions and (if R-parity is conserved) providing a plausible candidate for
cosmological dark matter (DM), supersymmetry also provides a framework
for the unification of particle physics and gravity. The latter is governed
by the Planck scale, $M_{P} \simeq10^{19}$ GeV, which is close to the
energy where the gravitational
interactions may become comparable in magnitude to the gauge interactions.

If supersymmetry is to be combined with gravity, it must be a
local symmetry, and the resulting theory is called supergravity
(SUGRA)~\cite{vNF,Deser77,Chamseddine82,Nilles84}. In models of
spontaneously-broken supergravity, the spin-3/2 superpartner of
the graviton, namely the gravitino $\widetilde{G}$, acquires a
mass $m_{\widetilde{G}}$ by absorbing the goldstino fermion of
spontaneously-broken supersymmetry via a super-Higgs
mechanism~\cite{Polonyi77,gangofmany}. The gravitino is a very
plausible candidate for the DM, since the presence of a massive
gravitino is an inevitable prediction of SUGRA. Moreover, in
models of gauge-mediated SUSY breaking~\cite{Fayet77} the
gravitino is necessarily the LSP. Furthermore, there are also
generic regions of the parameter space of minimal supergravity
(mSUGRA) models where the gravitino is the LSP.

In mSUGRA the soft supersymmetry-breaking parameters take a particularly
simple form at the Planck scale \cite{Nilles84}, in which the soft supersymmetry-breaking
scalar masses-squared $m_0^2$, gaugino masses $m_{1/2}$ and
trilinear parameters $A_0$ are each flavor diagonal and universal \cite{Hall83}. Renormalization-group
evolution (RGE) is then used to derive the effective values of the supersymmetric
parameters at the low-energy (electroweak) scale. Five parameters
fixed at the gauge coupling unification scale, $m_{0}$, the gaugino mass $m_{1/2}$, $A_{0}$,
$\tan\beta$ and the sign of $(\mu)$ are then related to the mass parameters at the scale
of electroweak symmetry breaking by the RGE. Moreover, mSUGRA requires that the
gravitino mass $m_{\widetilde G} = m_0$ at the Planck scale, and imposes a relation between
trilinear and bilinear soft supersymmetry-breaking parameters that may be used to fix $\tan \beta$
in terms of the other mSUGRA parameters. The mSUGRA parameter space offers both neutralino
dark matter and gravitino dark matter as generic possibilities.

In general, it would be difficult to detect gravitino DM directly, since the gravitino's couplings
are suppressed by inverse powers of the Planck scale. However, evidence for gravitino
DM may be obtained indirectly in collider experiments, since in such a case the next-to-lightest
sparticle (NLSP) has a long lifetime, as it decays into the gravitino through
very weak interactions suppressed by the Planck scale. One of the
most plausible candidates for the NLSP is the lighter scalar superpartner of the tau
lepton,  called the stau, $\widetilde{\tau}_1$ . If universal supersymmetry-breaking
soft supersymmetry-breaking masses for squarks and sleptons are assumed, as in
mSUGRA, renormalization effects and mixing between the
spartners of the left- and right-handed sleptons cause the $\widetilde{\tau}_{1}$ to
become the NLSP.

If R-parity is conserved, sparticles must be
pair-produced at colliders with sufficiently high energies, and heavier
sparticles will decay into other sparticles, until the decay cascade terminates
with the NLSP, which is metastable in gravitino DM scenarios based on mSUGRA
and decays only long after being produced.
Collider experiments offer possibilities to detect and measure metastable NLSPs
such as the staus in mSUGRA, and to use their decays to study
indirectly the properties of gravitinos,
which cannot be observed directly in astrophysical experiments.

There are several experimental constraints on the mSUGRA parameter
space: the LSP must be electrically and colour-neutral, and one
must satisfy the limit from LEP~2 on the stau mass:
$m_{\widetilde{\tau}}>85.2$ GeV obtained from the pair production
of staus, as well as limits on other sparticles, the mass of the
lightest Higgs scalar, the DM density, the branching ratio for
radiative \emph{B} decay and the muon anomalous magnetic moment.
An up-to-date compilation of these experimental and
phenomenological constraints on the mSUGRA model can be found
in~\cite{Allanach02,Baer02}.

Astrophysics and cosmology also constrain the properties of
metastable particles such as the staus in gravitino DM scenarios~\cite{Steffen06}, in
particular via the comparison between the calculated and observed abundances
of light elements, assuming the cosmological baryon density derived from
observations of the cosmic microwave background (CMB). In addition to the
hadronic and electromagnetic effects due to secondary particles produced in the
showers induced by stau decay, one must also consider the catalysis effects of
stau bound states~\cite{Pospelov}. In certain regions of mSUGRA parameter space, these may even
bring the calculations of Big-Bang nucleosynthesis (BBN) into better agreement with
the observed abundances of $^{6,7}$Li~\cite{Cyburt06}. A number of
benchmark scenarios in which the stau decay showers do not destroy the
successful results of BBN were proposed in~\cite{Bench3}. These did
not take bound-state effects into account, but these may be included in related models,
as we discuss later.

The phenomenology of stau NLSPs in both proton-proton and $e^{+}e^{-}$
colliders has been considered in a number of previous papers. In particular,
it has been shown that the stau mass could be measured quite accurately by the
ATLAS and CMS experiments at the LHC, using their tracking systems and time-of-flight
(ToF) information from Resistive Plate Chamber (RPCs).
Previous studies show that the polarization of $\tau$ lepton
could be measured via the energy distribution of the decay products
of the polarized $\tau$ lepton at future lepton colliders~\cite{Nojiri95}.
There have also been some phenomenological investigations of long-lived staus
at the ILC~\cite{Allanach02,Feng05,Khotilovich05,Brandenburg05,Martyn06}.

In this work, we study the pair production of staus in $e^{+}e^{-}$
annihilation at the future linear colliders ILC~\cite{ILC} and CLIC~\cite{CLIC},
assuming that the LSP is the gravitino and the NLSP is the lighter stau.
In addition to the three benchmark scenarios proposed earlier, we also propose a
new benchmark scenario chosen to improve the agreement of Big-Bang
nucleosynthesis calculations with the observed $^{6,7}$Li abundances after bound-state
catalysis effects are included. This point
has a relatively heavy stau that could only be observed at CLIC. Following a
general discussion of the signatures and detection of staus in the linear-collider
environment, we explore the possible measurements of model parameters such as
the stau mixing angle that could be made at the ILC or CLIC. We show that beam
polarization and/or a measurement of the final-state $\tau$ polarization would be very
useful in this respect.
 We also discuss the
choice of the center-of-mass energy that would maximize the cross section for
producing slow-moving staus with $\beta \gamma < 0.4$, that would be trapped in a
typical linear-collider detector. The subsequent stau decays would be optimal for some
measurements of gravitino parameters. We show that most such slow-moving staus
are produced in the cascade decays of heavier sparticles, so that the optimal
center-of-mass energies for trapping staus are typically considerably higher than the
stau pair-production threshold.

\section{The Model Parameters}

In the mSUGRA framework used here, $\mu$ is
the supersymmetric Higgs(ino) mass parameter, we denote by $m_{0}$ the universal soft
SUSY-breaking contribution to the masses of all scalars (including the Higgs
fields), $m_{1/2}$ is the universal SUSY-breaking
gaugino mass, and $A_{0}$ is the universal SUSY-breaking factor in the
soft trilinear scalar
interactions. In addition, there is a bilinear soft SUSY-breaking parameter $B_0$,
which appears in the effective Higgs potential, and is related by mSUGRA to the
trilinear parameter: $B_0 = A_0 - m_0$. All these
soft SUSY-breaking parameters are defined at the scale of grand unification. In
general studies of the MSSM, $\tan \beta$, the ratio of the vacuum expectation
values (vevs) of the two Higgs doublets at the weak scale, is taken as a free
parameter. However, in mSUGRA it is determined dynamically once the relation
between $A_0$ and $B_0$ is imposed. This parameter
set also determines the gravitino mass, since $m_{\widetilde G} = m_0$ in mSUGRA,
as well as the other sparticle masses.
For calculations of sparticle mass spectra and other properties in the models we study here,
we use the updated {\tt ISASUGRA}
programme, which is part of the {\tt ISAJET} package~\cite{Baer00}.
This programme is convenient for event generation with {\tt PYTHIA}~\cite{Sjostrand06}
for various colliders.

A selection of benchmark mSUGRA points consistent
with present data from particle physics and BBN constraints on the decays of
metastable NLSPs was proposed in~\cite{Cyburt06}. We use a subset
of these benchmark scenarios to study mSUGRA phenomenology at $e^+ e^-$ colliders,
namely the points denoted $(\epsilon,\zeta,\eta)$, in which the NLSP is the lighter stau,
$\tilde{\tau}_{1}$. The $\tilde{\tau}_{1}$ lifetime in these models ranges from
$\sim 10^4$~s to $\sim 3\times10^{6}$~s, so fast-moving charged
NLSPs would be indistinguishable from massive stable particles, in a generic collider
detector. However,
it might be possible to observe the decays into gravitinos of any
slow-moving $\tilde{\tau}_{1}$s that
lose sufficient energy to stop in the detector.

\begin{figure}
\includegraphics[scale=0.45]{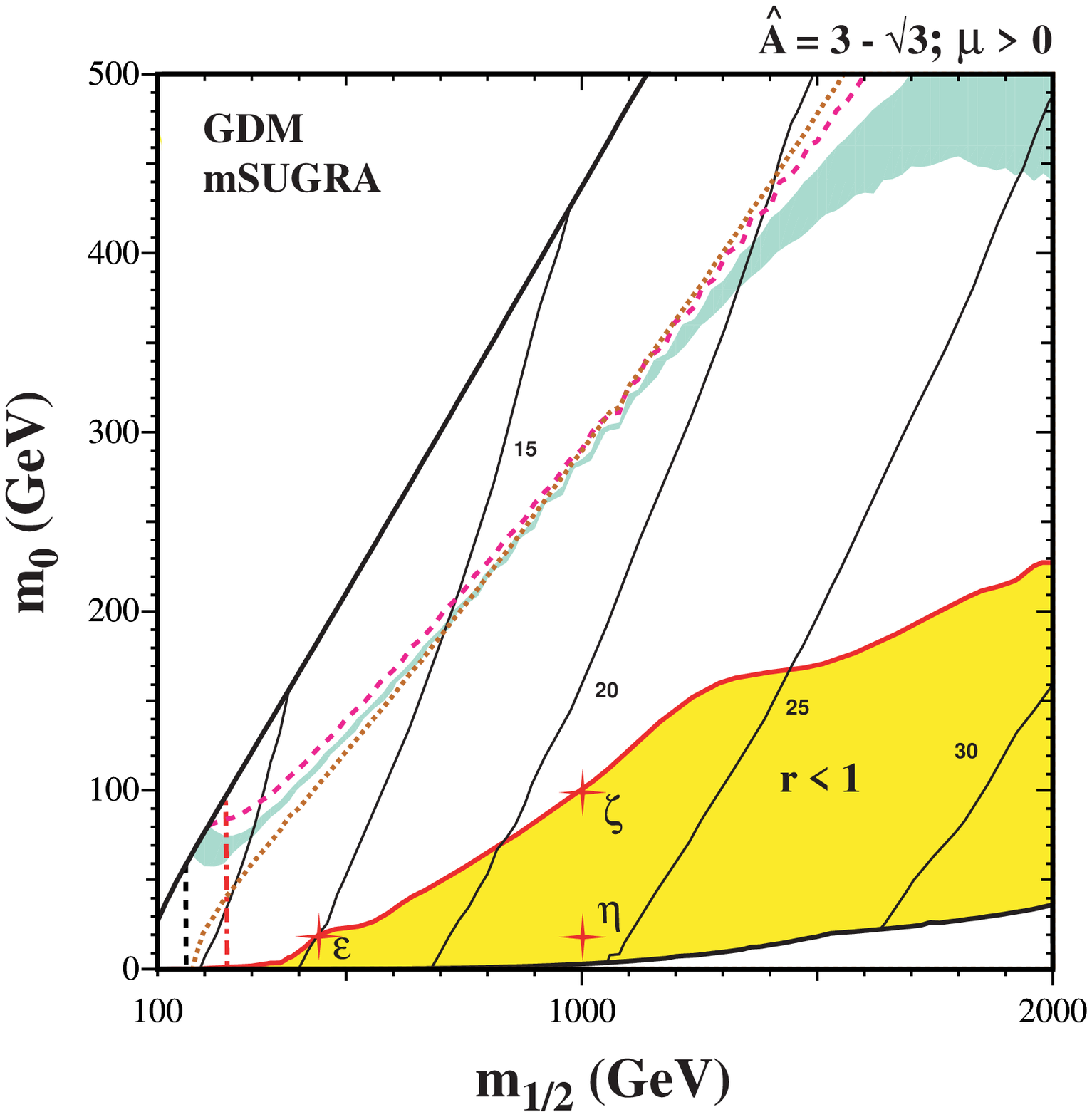}
\includegraphics[scale=0.45]{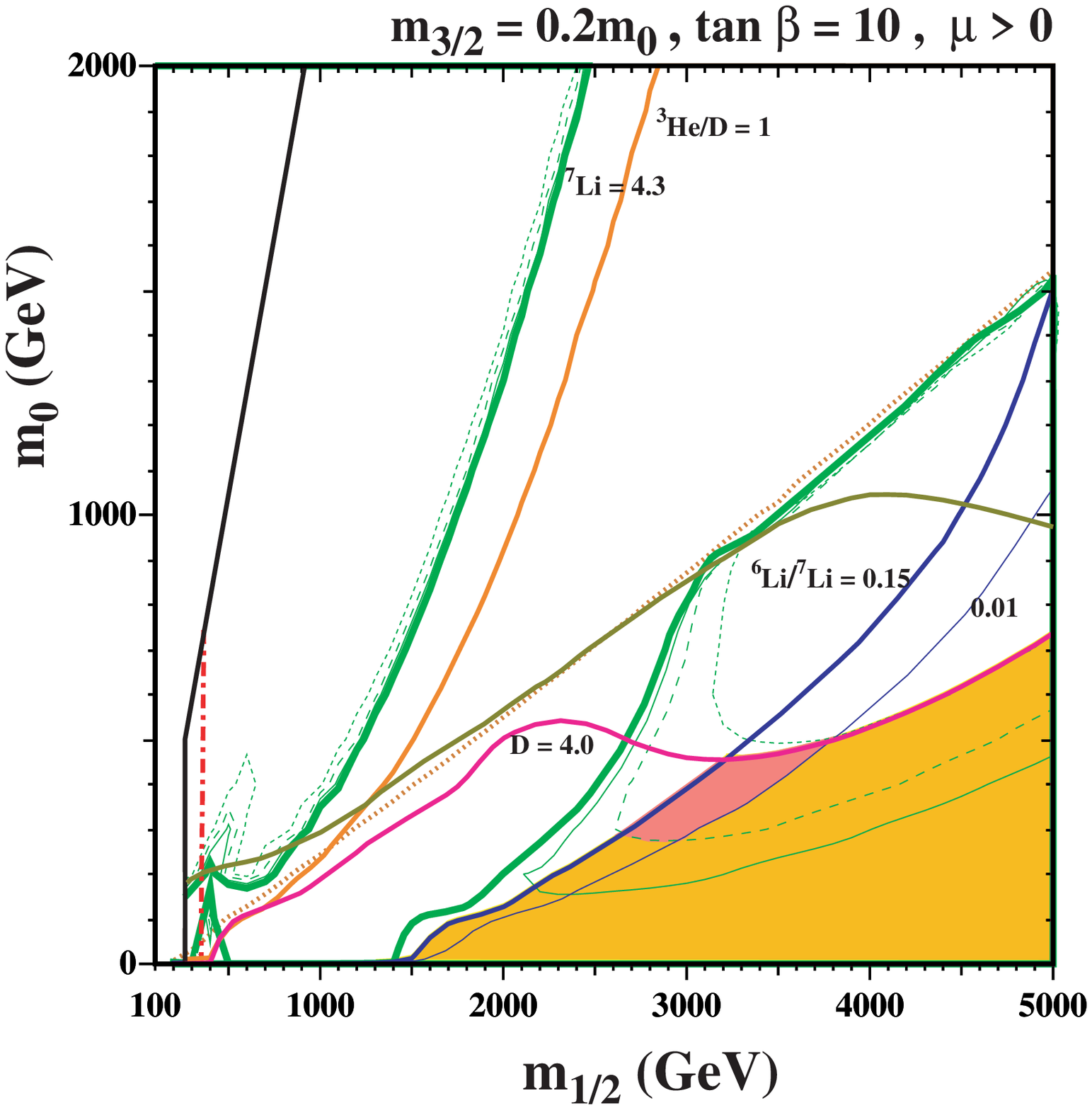}
\caption{\it The left (first) panel shows the $(m_{0},m_{1/2})$ plane for mSUGRA with
a gravitino LSP and stau NLSP, assuming $A_0 = 3 - \sqrt{3}$, with the regions allowed by
astrophysical constraints on stau decays~\cite{Bench3} shaded light (yellow). The
values of $\tan\beta$ are fixed by the vacuum conditions with the values indicated along
the solid black contours, and the three benchmark points $\epsilon$, $\eta$ and $\zeta$ are
shown by red $+$ signs. The right (second) panel shows the $(m_{0},m_{1/2})$ plane in the CMSSM,
assuming fixed $\tan \beta = 10$, with the regions allowed in the presence of
cosmological bound-state effects and stau decays if $m_{\widetilde G} = 0.2 m_0$
shaded light (yellow). In the darker (pink) shaded region, bound-state effects yield $^{6,7}$Li
abundances in the ranges favoured by astrophysical observations. The new benchmark
point $\theta$ is located in this $^{6,7}$Li-friendly region~\cite{Cyburt06}.
\label{fig1}}
\end{figure}

The mSUGRA benchmark scenarios were formulated assuming the
representative value $A_{0}=(3-\sqrt{3})m_{0}$ found in the simplest Polonyi
model of supersymmetry breaking in mSUGRA~\cite{Polonyi77}.
The allowed region of the gravitino DM parameter space may be displayed
in the $(m_{1/2},m_{0})$ plane, as shown in the left panel
of Fig.~\ref{fig1} for the Polonyi value of $A_0$. The theoretical and
phenomenological constraints are displayed, together with the astrophysical
constraints on NLSP decays. The combined effect of these constraints is to
allow a wedge in the $(m_{1/2},m_{0})$ plane at relatively low values of $m_0$.
The benchmark point $\epsilon$ was chosen near the apex of this wedge,
whereas the points $\eta$ and $\zeta$ were chosen at larger values of $m_{1/2}$,
and hence more challenging for the LHC and other colliders~\cite{Bench3}.
The mSUGRA parameters
specifying these models and some of the corresponding sparticle masses are shown
in Table~\ref{table1}, as calculated using
{\tt ISAJET 7.74}~$^1$~\footnotetext[1]{In addition, we note that the ${\tilde e}_R - {\tilde \tau}_1$
mass difference for point $\theta$ is just $O$(100~MeV).}.

\begin{table}
\caption{\it The four benchmark points with gravitino DM and a
$\tilde{\tau}_{1}$ NLSP that are considered in this paper. The
quoted slepton, squark, gluino and the neutralino mass spectra (in
GeV) have been calculated with {\tt ISAJET 7.74}, using
$m_{t}=175$~GeV. \label{table1}}
\begin{center}
\begin{tabular}{|c|c|c|c|c|}
\hline
Model &
 $\epsilon$&
 $\zeta$&
 $\eta$&
$\theta$
\tabularnewline
\hline
 $m_{0}$&
20&
100&
20&
330\tabularnewline
\hline
$m_{1/2}$&
 440 &
 1000 &
 1000 &
3000\tabularnewline
\hline
 $A_{0}$&
 -25 &
 -127 &
-25 &
0\tabularnewline
\hline
 tan$\beta$&
15 &
 21.5 &
 23.7 &
10\tabularnewline
\hline
 sign($\mu$) &
 +1&
 +1&
 +1&
 +1
\tabularnewline
\hline
$\tilde{e}_{L},\tilde{\mu}_{L}$&
 303 &
 676 &
 669 &
1982\tabularnewline
\hline
 $\tilde{e}_{R},\tilde{\mu}_{R}$&
 168 &
 382 &
 369 &
1140\tabularnewline
\hline
 $\tilde{\nu}_{e},\tilde{\nu}_{\mu}$&
 289 &
 666 &
 659 &
1968\tabularnewline
\hline
 $\tilde{\tau}_{1}$&
 154 &
 346 &
 327 &
1140\tabularnewline
\hline
 $\tilde{\tau}_{2}$&
 304 &
 666 &
 659 &
1966\tabularnewline
\hline
 $\tilde{\nu}_{\tau}$&
 284 &
 651 &
 643 &
1944\tabularnewline
\hline
$\tilde{u}_L,\tilde{c}_L$&
935 & 1992 & 1989 &
5499\tabularnewline
\hline
$\tilde{u}_R,\tilde{c}_R$&
902 & 1913 & 1910 &
5248\tabularnewline
\hline
$\tilde{d}_L,\tilde{s}_L$&
938 & 1994 & 1991 &
5500\tabularnewline
\hline
$\tilde{d}_R,\tilde{s}_R$&
899 & 1903 & 1900 &
5217\tabularnewline
\hline
$\tilde{t}_1$&
703 & 1534 & 1541 &
4285\tabularnewline
\hline
$\tilde{t}_2$&
908 & 1857 & 1855 &
5130\tabularnewline
\hline
$\tilde{b}_1$&
858 & 1823 & 1819 &
5104\tabularnewline
\hline
$\tilde{b}_2$&
894 & 1874 & 1867 &
5203\tabularnewline
\hline
$\tilde{g}$&
1023 & 2187 & 2186 &
6089\tabularnewline
\hline
$\tilde{\chi}_1^0$&
179 &  425 & 424 &
1336 \tabularnewline
\hline
$\tilde{\chi}_2^0$&
337 &  802 & 802 &
2467 \tabularnewline
\hline
$\tilde{\chi}_1^\pm$&
338 &  804 & 804 &
2472 \tabularnewline
\hline
\end{tabular}
\end{center}
\end{table}

Recently, bound-state effects on light-element abundances in models with
metastable stau NLSPs have been studied in several variants of the
MSSM~\cite{Cyburt06,Pradler06},
including mSUGRA and some other models in which its relations between
$m_{\tilde{G}}$ and $m_0$ and between $A_0$ and $B_0$ are relaxed, but
where universality is maintained for the soft SUSY-breaking parameters, a
framework we term the constrained MSSM (CMSSM).
The bound-state effects restrict greatly the allowed regions of parameter
space, but one example was found within the CMSSM of an $(m_{1/2}, m_0)$ plane with
$m_{\tilde{G}} = 0.2 m_0$ and $\tan \beta = 10$,
that has a wedge compatible with BBN and also a small
region where the BBN predictions for the $^{6,7}$Li abundances could be
brought into better agreement with astrophysical observations.
We propose and study here a new benchmark point in this $^{6,7}$Li-friendly region,
denoted by $\theta$, with the parameter choices shown in the last
column of Table~\ref{table1}. Also shown in this column are some sparticle masses:
like the rest of the $^{6,7}$Li-friendly region, benchmark $\theta$ features
relatively heavy sparticles beyond the reach of either the LHC or the ILC,
but within the kinematic reach of CLIC.

In this connection we note that for benchmark $\theta$ the direct
stau pair production cross section is $2.7\times10^{-6}$ pb at the LHC with $\sqrt{s} = 14$~TeV
and $2.8\times10^{-5}$ pb at the DLHC with $\sqrt{s} = 28$~TeV. The plausible integrated luminosity
of $10^{5}$ pb$^{-1}$ for the LHC would be insufficient to reach this point, and it would be very
marginal at the SLHC with $10^{6}$ pb$^{-1}$. The only alternative to CLIC would be the DLHC with
a very high integrated luminosity.

The bound-state effects depend sensitively on the stau lifetime, being much
more important for longer lifetimes. This is why they would
be unacceptable for the the stau in benchmark scenarios
$\epsilon$, $\eta$ and $\zeta$, if it had the lifetime predicted by mSUGRA.
However, as discussed in the following Section,
the stau lifetime depends sensitively on the mass assumed for
the gravitino. Hence, a CMSSM variant with similar masses for the masses of the spartners
of the Standard Model particles but a smaller gravitino mass would not be excluded
{\it a priori} on the basis of bound-state effects. For this reason, and since the
previous benchmarks $\epsilon$, $\eta$ and $\zeta$ represent to a great
extent the range of possibilities open to the ILC (limits on metastable particles
exclude a stau much lighter than at benchmark $\epsilon$, and sparticles
would be inaccessible to the LHC or ILC if they weighed much more than at
points $\eta$ and $\zeta$), we continue to include all these points in the
subsequent discussion, though our primary interest will be in the new
$^{6,7}$Li-friendly point $\theta$.

\section{The $\widetilde{\tau}_{1}$ NLSP Decay and Lifetime}

Since we assume the conservation of R-parity, the gravitino must be
produced at the end of any decay chain initiated by the decay of a heavy
unstable supersymmetric particle. As already remarked, in models with
universal scalar soft supersymmetry-breaking masses, the lightest slepton
is generically the lighter stau mass eigenstate, the $\widetilde{\tau}_{1}$, as
a result of renormalization-group effects and left-right sfermion mixing.
In large regions of mSUGRA parameter space, the $\widetilde{\tau}_{1}$ is
the NLSP, and is the penultimate sparticle in (essentially) every decay chain.

The $\widetilde{\tau}_{1}$
is in general a linear combination of $\widetilde{\tau}_{L}$ and
$\widetilde{\tau}_{R}$, which are the superpartners of the left-
and right-handed $\tau$ leptons $\tau_{L}$ and $\tau_{R}$, respectively.
In general the stau mass eigenstates are
\begin{align}
\widetilde{\tau}_{1} & =\widetilde{\tau}_{L}\cos\theta_{\tilde{\tau}}+\widetilde{\tau}_{R}\sin\theta_{\tilde{\tau}},\\
\widetilde{\tau}_{2} & =-\widetilde{\tau}_{L}\sin\theta_{\tilde{\tau}}+\widetilde{\tau}_{R}\cos\theta_{\tilde{\tau}},\end{align}
 where $\theta_{\tilde{\tau}}$ is the stau mixing angle, which is given by
\begin{equation}
\cos\theta_{\tilde{\tau}}=\frac{-m_{\tau}(A_{\tau}-\mu\tan\beta)}{\sqrt{(m_{\tilde{\tau}_{L}}^{2}-m_{\tilde{\tau}_{1}}^{2})+m_{\tau}^{2}(A_{\tau}-\mu\tan\beta)^{2}}},
\end{equation}
where $|\cos\theta_{\tilde{\tau}}|>1/\sqrt{2}$ if $m_{\widetilde{\tau}_{L}}<m_{\widetilde{\tau}_{R}}$
and $|\cos\theta_{\tilde{\tau}}|<1/\sqrt{2}$ if $m_{\widetilde{\tau}_{R}}<m_{\widetilde{\tau}_{L}}$.

The interactions of the stau states $\widetilde{\tau}_{L,R}$ with the gravitino
$\widetilde{G}$ and tau lepton $\tau$ are described by the Lagrangian~\cite{Buchmuller04}
\begin{eqnarray}
L_{\widetilde{\tau}\tau\tilde{G}} & = & -\frac{1}{\sqrt{2}M_{P}}\left[(D_{\nu}\widetilde{\tau}_{L})^{*}
\overline{\widetilde{G}}^{\mu}\gamma^{\nu}\gamma_{\mu}P_{L}\tau+(D_{\nu}\widetilde{\tau}_{L})
\overline{\tau}P_{L}\gamma_{\mu}\gamma^{\nu}\widetilde{G}^{\mu}\right. \nonumber \\
 &  & \left.+(D_{\nu}\widetilde{\tau}_{R})^{*}\overline{\widetilde{G}}^{\mu}
 \gamma^{\nu}\gamma_{\mu}P_{R}\tau+(D_{\nu}\widetilde{\tau}_{R})
 \overline{\tau}P_{R}\gamma_{\mu}\gamma^{\nu}\widetilde{G}^{\mu}\right],
\end{eqnarray}
where $D_{\nu} \equiv (\partial_{\nu}+ig_{e}A_{\nu})$, with $A_{\nu}$ denoting
the gauge boson and $M_{P} \equiv (8\pi G_{N})^{-1/2}=2.436\times10^{18}$
GeV is the reduced Planck mass, with the Newton constant $G_{N}=6.707\times10^{-39}$
GeV$^{-2}$.

\begin{figure}
\includegraphics{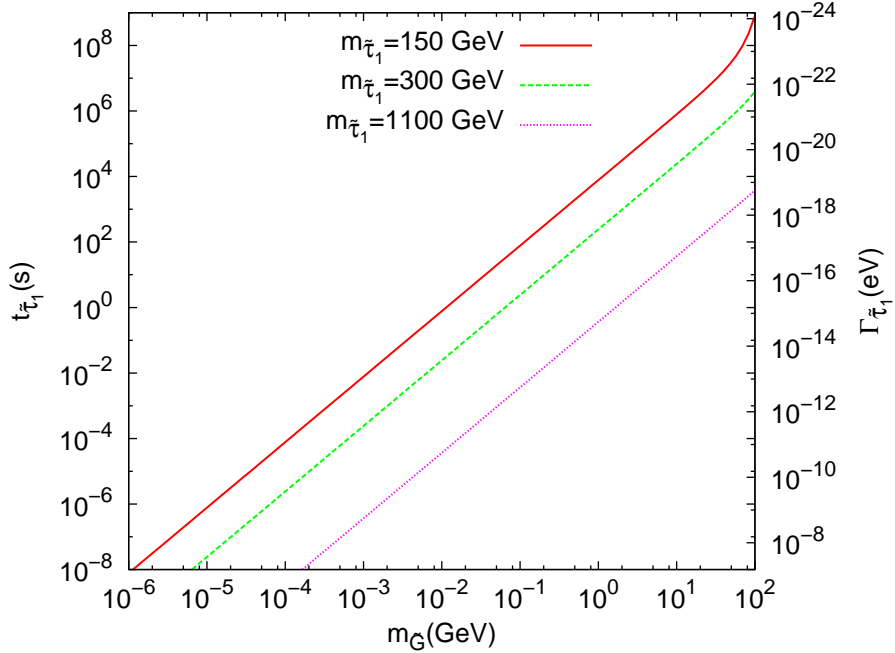}
\caption{\it The stau NLSP lifetime and decay width as functions of the gravitino mass, for
various choices of stau mass.
\label{fig2}}
\end{figure}

The stau decay rate is dominated by the two-body decay into tau and
gravitino ($\widetilde{\tau}_{1}\rightarrow\tau\widetilde{G}$), and the
decay width of the $\widetilde{\tau}_{1}$ is given by
\begin{equation}
\Gamma_{\widetilde{\tau_{1}}}=\frac{(m_{\widetilde{\tau}_{1}}^{2}-m_{\widetilde{G}}^{2}-m_{\tau}^{2})^{4}}{48\pi M_{p}^{2}m_{\widetilde{G}}^{2}m_{\widetilde{\tau}_{1}}^{3}}\left(1-\frac{4m_{\widetilde{G}}^{2}m_{\tau}^{2}}{(m_{\widetilde{\tau}_{1}}^{2}-m_{\widetilde{G}}^{2}-m_{\tau}^{2})^{2}}\right)^{3/2},
\end{equation}
where $m_{\widetilde{\tau}_{1}}$, $m_{\widetilde{G}}$ and $m_{\tau}$
are the masses of the stau $\widetilde{\tau}_{1}$, gravitino $\widetilde{G}$
and tau lepton $\tau$, respectively. As illustrative examples, if we take
$m_{\widetilde{\tau}_{1}}=150$~GeV (similar to the mass at benchmark point $\epsilon$),
we find a lifetime of $\Gamma_{\widetilde{\tau_{1}}}^{-1}=7762$~s (25.9 years) for a gravitino
mass of $m_{\widetilde{G}}=1$ GeV (100~GeV),
respectively. The lifetime and decay width of the stau as functions of the gravitino
mass for various stau mass values are presented in Fig.~\ref{fig2}.

\begin{figure}
\includegraphics{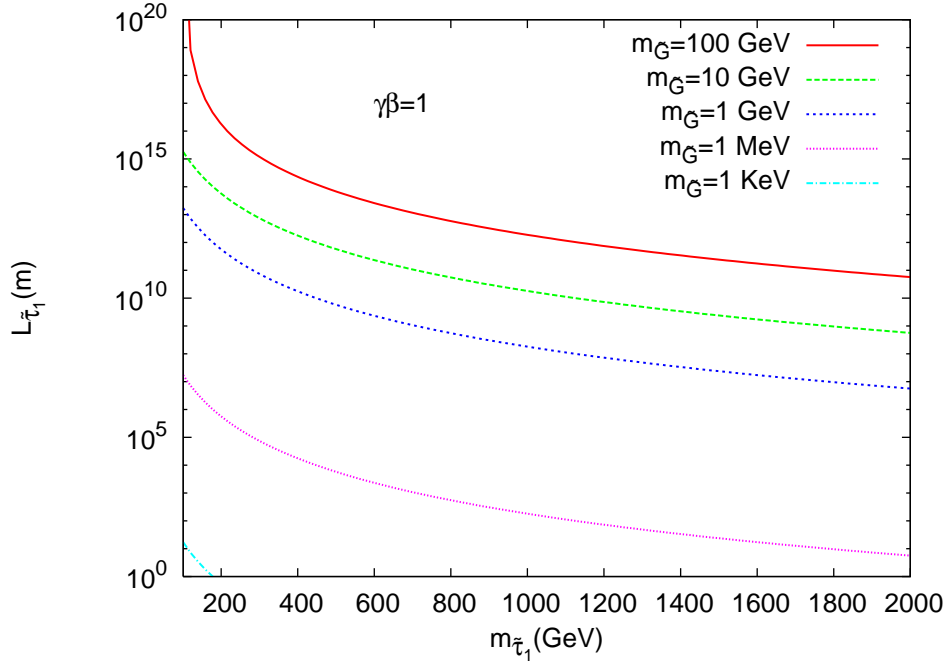}
\caption{\it The mean decay length of the stau NLSP as a function of its mass, for various
choices of the gravitino mass.
\label{fig3}}
\end{figure}

Neglecting the tau lepton mass,
the mean stau decay length, obtained from the relation $L=\beta\gamma t_{\tilde{\tau}}$, is
shown in Fig.~\ref{fig3}. For instance, taking $\beta\gamma=1$ and
$m_{\widetilde{G}}=1$~MeV would imply a mean decay length of $L=2.33\times10^{6}$~m
($L=109.6$~m) for $m_{\tilde{\tau_1}}=150$~GeV
($m_{\tilde{\tau_1}}=1100$~GeV), respectively. Only for $m_{\widetilde{G}} < 1$~KeV
might a typical stau with $\beta \gamma = {\cal O}(1)$ decay within a generic collider
detector. In this case the light gravitino
would constitute either warm or even hot dark matter, which is disfavoured by the
modelling of cosmological structure formation.

However, some fraction of the staus may be produced with
velocities sufficiently low that strong ionization energy loss
causes them to stop within the detector. In such a case, interesting
information may be extracted from the stau decays. For example,
the mass of the gravitino produced in the decay could in principle
be inferred kinematically from the relation
$m_{\widetilde{G}}^{2}=
m_{\tilde{\tau}_{1}}^{2}+m_{\tau}^{2}-2m_{\tau}E_{\tau}$, unless
it is very small. Also, we note that the polarization of the tau
produced in the stau decay must coincide with the combination of
${\tilde \tau_{L,R}}$ contained in the ${\tilde \tau_1}$, offering
in principle the possibility of determining the stau mixing angle
$\theta_{\tilde{\tau}}$ by examining the kinematical distributions
in the subsequent tau decays, as well as in the polarization dependence of stau
pair production. We discuss later the prospects for measuring
$\theta_{\tilde{\tau}}$ using the stau pair-production cross
section in $e^+ e^-$ collisions.

The point $\epsilon$ is near the tip of the wedge in Fig.~\ref{fig1}, and
the stau NLSP $\tilde{\tau}_{1}$ has a mass of
$m_{\tilde{\tau}}=154$ GeV and a lifetime of
$t_{\tilde{\tau}}=2.9\times10^{6}$ s in
mSUGRA~$^2$~\footnotetext[2]{We re-emphasize that the stau lifetimes
at the other benchmark points would be reduced for smaller
gravitino masses, as might be preferred by the observed
light-element abundances.}. The benchmark point $\zeta$ is close
to the upper edge of the wedge, with $m_{0}=100$ GeV and
$\tan\beta\simeq21.5$, and the $\tilde{\tau}_{1}$ has a mass of
$346$~GeV and a lifetime of $1.7\times10^{6}$~s in mSUGRA. The benchmark
point $\eta$ is close to the lower edge of the wedge in Fig.~\ref{fig1},
with $m_{0}=20$ GeV and $\tan\beta=23.7$. Here the
$\tilde{\tau}_{1}$ has a mass of $m_{\tilde{\tau}}=327$~GeV and a
smaller lifetime of $6.4\times10^{4}$~s in mSUGRA. The new benchmark point
$\theta$ introduced here with a much larger value of $m_{1/2}$,
corresponding to $m_{\tilde{\tau}}=1140$ GeV and
$t_{\tilde{\tau}}=1.4\times10^{3}$~s. The stau mixing angle
$\theta_{\tilde{\tau}}$ takes the following values at these
points: $77.2^{o}$, $82.4^{o}$, $81.8^{o}$ and $89.0^{o}$ for the
points $\epsilon$, $\zeta$, $\eta$ and $\theta$, respectively.
Thus, each of these scenarios predicts that the ${\tilde \tau_1}$
should be mainly {\it right-handed}. Then, the {\it right-handed}
polarization of the electron beam enhances the $\stau_1^+\stau_1^-$ signal.
We expect this feature to be
quite general, and more marked for heavier staus. Correspondingly,
the tau produced in stau decay should also be mainly {\it
right-handed}.

\section{Production of Stau Pairs in $e^+ e^-$ Annihilation}

\begin{figure}
\includegraphics{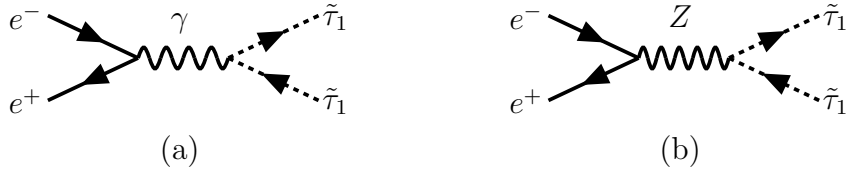}
\caption{\it The tree-level Feynman diagrams for the reaction
$e^{+}e^{-}\rightarrow\tilde{\tau}_{1}^{+}\tilde{\tau}_{1}^{-}$.
\label{fig4}}
\end{figure}

The reaction $e^{+}e^{-}\rightarrow\tilde{\tau}_{1}^{+}\tilde{\tau}_{1}^{-}$
proceeds via direct-channel $\gamma$ and $Z$ exchange as shown in Fig. \ref{fig4}.
The tree-level vertex factors can be parametrized as given in Table~\ref{table1} for
the stau-photon-stau ($\widetilde{\tau}_{1}^{+}\gamma\tilde{\tau}_{1}^{-}$)
and stau-Z-stau ($\widetilde{\tau}_{1}^{+}Z\tilde{\tau}_{1}^{-}$)
interactions.

\begin{table}

\caption{\it The vertices for
$\widetilde{\tau}_{1}\tau\widetilde{G}$,
$\widetilde{\tau}_{1}^{+}\tilde{\tau}_{1}^{-}\gamma$ and
$\widetilde{\tau}_{1}^{+}\tilde{\tau}_{1}^{-}Z$ interactions and
the corresponding tree-level diagrams. The parameters $C_{L,R}$
depend on the stau mixing angle $\theta_{\tilde \tau}$ and the
coefficients $C_{1,2}$ depend also on the weak mixing angle. $F$
is the supersymmetry breaking scale.\label{table2}}

\begin{center}\begin{tabular}{llll}
\hline
 \multicolumn{1}{c}{}
 \includegraphics[
  width=3cm,
  height=1.5cm]{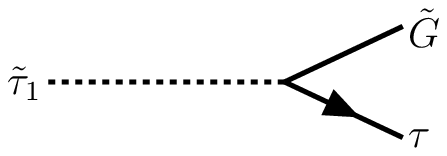}&
  &
 $\frac{1}{2F}\left[(\not p_{2}\not p_{3}+\not p_{3}\not p_{2})[C_{L}(1-\gamma_{5})+C_{R}(1+\gamma_{5}]\right]$ \tabularnewline
 \multicolumn{1}{c}{}&
 &
 \multicolumn{1}{c}{}\tabularnewline
 \includegraphics[
  width=3cm,
  height=1.5cm]{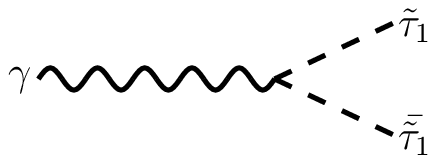}&
&
 $ig_{e}(p_{3\mu}-p_{2\mu})$\tabularnewline
\multicolumn{1}{c}{}&
&
\multicolumn{1}{c}{}\tabularnewline
\includegraphics[
  width=3cm,
  height=1.5cm]{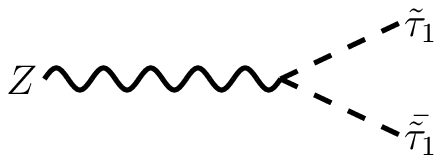}&
&
 $-\frac{ig_{e}}{\sin2\theta_{W}}(C_{1}p_{3\mu}+C_{2}p_{2\mu})$\tabularnewline
\multicolumn{1}{c}{}&
&
\multicolumn{1}{c}{}\tabularnewline

\hline
\end{tabular}
\end{center}
\end{table}

The cross section for the process $e^+e^-\to\tilde{\tau}_{1}^+\tilde{\tau}_1^-$
with a left (L)- or right (R)-polarized $e^\pm$ beams at a center-of-mass
energy $\sqrt{s}$ is given by \cite{kraml99}.
\begin{eqnarray}
\sigma&=&\frac{\pi\alpha^{2}\beta^{3}}{3s}
\left[(1-P_-P_+)+\frac{I_3\cos^2\theta_{\tilde{\tau}}
+\sin^2\theta_W}{2\cos^2\theta_W\sin^2\theta_W}
[v_e(1-P_-P_+)-a_e(P_--P_+)]P_{\gamma Z} \right. \nonumber \\
&+&\left.\frac{(I_3\cos^2\theta_{\tilde{\tau}}+\sin^2\theta_W)^2}{16\cos^4\theta_W\sin^4\theta_W}
[(v_e^2+a_e^2)(1-P_-P_+)-2v_ea_e(P_--P_+)]P_{ZZ}\right],\\ \nonumber
\label{sigma}
\end{eqnarray}
where $v_{e} = 4\sin^{2}\theta_{W}-1$ and $a_{e}=-1$. $P_-$ and $P_+$ denote the degree of polarization of the $e^-$ and $e^+$ beams, respectively. The propagator terms are given by
\begin{equation}
P_{ZZ}=\frac{s^2}{(s-m_{Z}^{2})^{2}+m_{Z}^{2}\Gamma_{Z}^{2}} \quad,\qquad
P_{\gamma Z}=\frac{s(s-m_Z^2)}{(s-m_{Z}^{2})^{2}+m_{Z}^{2}\Gamma_{Z}^{2}}
\end{equation}

\begin{figure}
\includegraphics{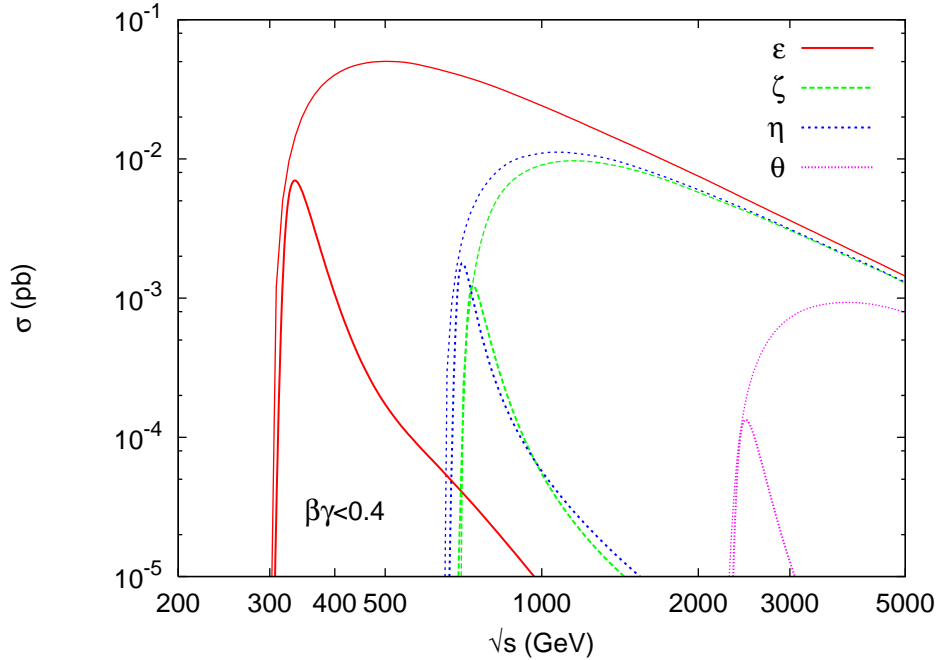}
\caption{\it The dependences of the tree-level cross sections for $\tilde{\tau}_{1}$ pair production
on the linear-collider center-of-mass energy, as calculated using {\tt PYTHIA} and
{\tt ISASUGRA}. We also show the cross sections for
producing slow staus with $\beta \gamma < 0.4$
in the various benchmark scenarios, also as
functions of the center-of-mass energy of the linear $e^{+}e^{-}$ collider.
The optimal energies for obtaining the
largest numbers of these stoppable staus are
different from the energies where the maxima of the pair-production cross sections appear.
\label{fig5}}
\end{figure}

The cross sections with unpolarized
electron and positron beams in the four benchmark scenarios,
shown in Fig.~\ref{fig5},
exhibit a typical $\beta^{3}$ threshold suppression, where
$\beta \equiv \sqrt{1-4m_{\tilde{\tau}}^{2}/s}$.
Illustrative values of the cross sections for producing $\tilde{\tau}_{1}$ pairs in
the benchmark scenarios $\epsilon,\zeta, \eta$, and $\theta$
are shown in Table~\ref{table3}, for
center-of-mass energies between 0.5 and 5~TeV. As an example,
with an integrated luminosity $L_{int}=200$ fb$^{-1}$ at
$\sqrt{s}=1$ TeV, the benchmark points $\epsilon,\zeta$ and $\eta$
would produce 4882, 1846 and 2150 $\tilde{\tau}_{1}$ pairs, respectively.
In general, the peak cross sections are found at center-of-mass
energies $\sqrt{s} \simeq 3 m_{\tilde \tau_1}$, namely $\sqrt{s} \simeq
500, 1000$ and 4000~GeV for benchmarks $\epsilon$,
$\eta$ and $\zeta$, and $\theta$, respectively. These would be good energies
for high-statistics studies of stau properties. The mass of the $\stau$ can
be estimated from the threshold scan and the mixing angle from the
polarized cross sections.

\begin{table}
\caption{\it The total cross section in pb calculated using {\tt PYTHIA} with the
full {\tt ISASUGRA} spectrum, including both initial- and final-state
radiation (ISR+FSR).
\label{table3}}
\begin{tabular}{cccccc}
\hline
&
Benchmark points&
$\epsilon$&
$\zeta$&
$\eta$&
$\theta$\tabularnewline
\hline
&
$m_{\tilde{\tau}_{1}}$(GeV) $=$&
154&
346&
327&
1140\tabularnewline
\hline
&
500&
$4.799\times10^{-2}$&
$-$&
$-$&
-\tabularnewline
$\sqrt{s}$(GeV)&
1000&
$2.441\times10^{-2}$&
$9.230\times10^{-3}$&
$1.075\times10^{-2}$&
-\tabularnewline
&
3000&
$3.665\times10^{-3}$&
$3.142\times10^{-3}$&
$3.197\times10^{-3}$&
$7.235\times10^{-4}$\tabularnewline
&
5000&
$1.432\times10^{-3}$&
$1.299\times10^{-3}$&
$1.311\times10^{-3}$&
$7.889\times10^{-4}$\tabularnewline
\hline
\end{tabular}
\end{table}

\begin{figure}
\includegraphics[
  scale=0.6]{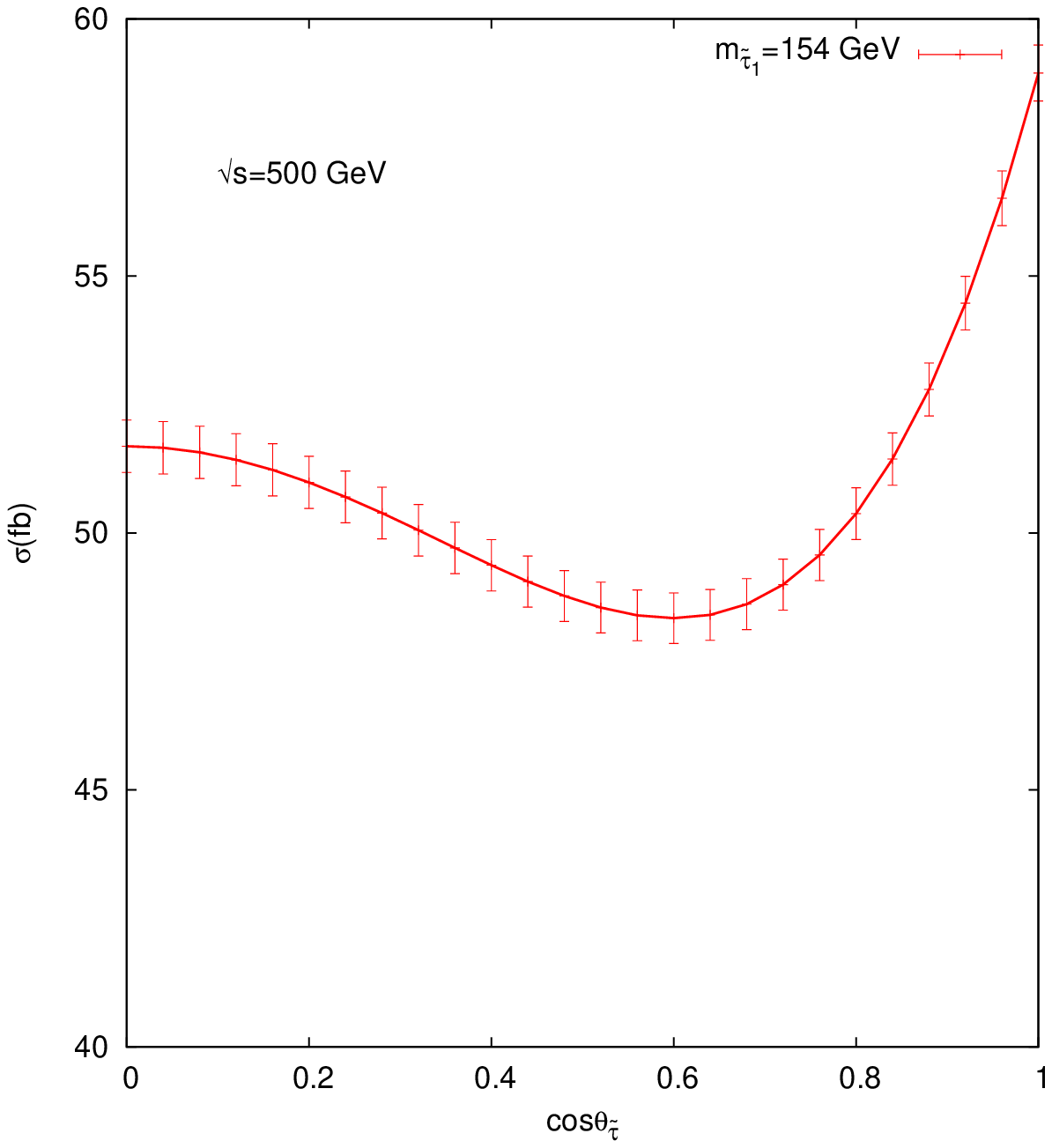}
\includegraphics[
  scale=0.6]{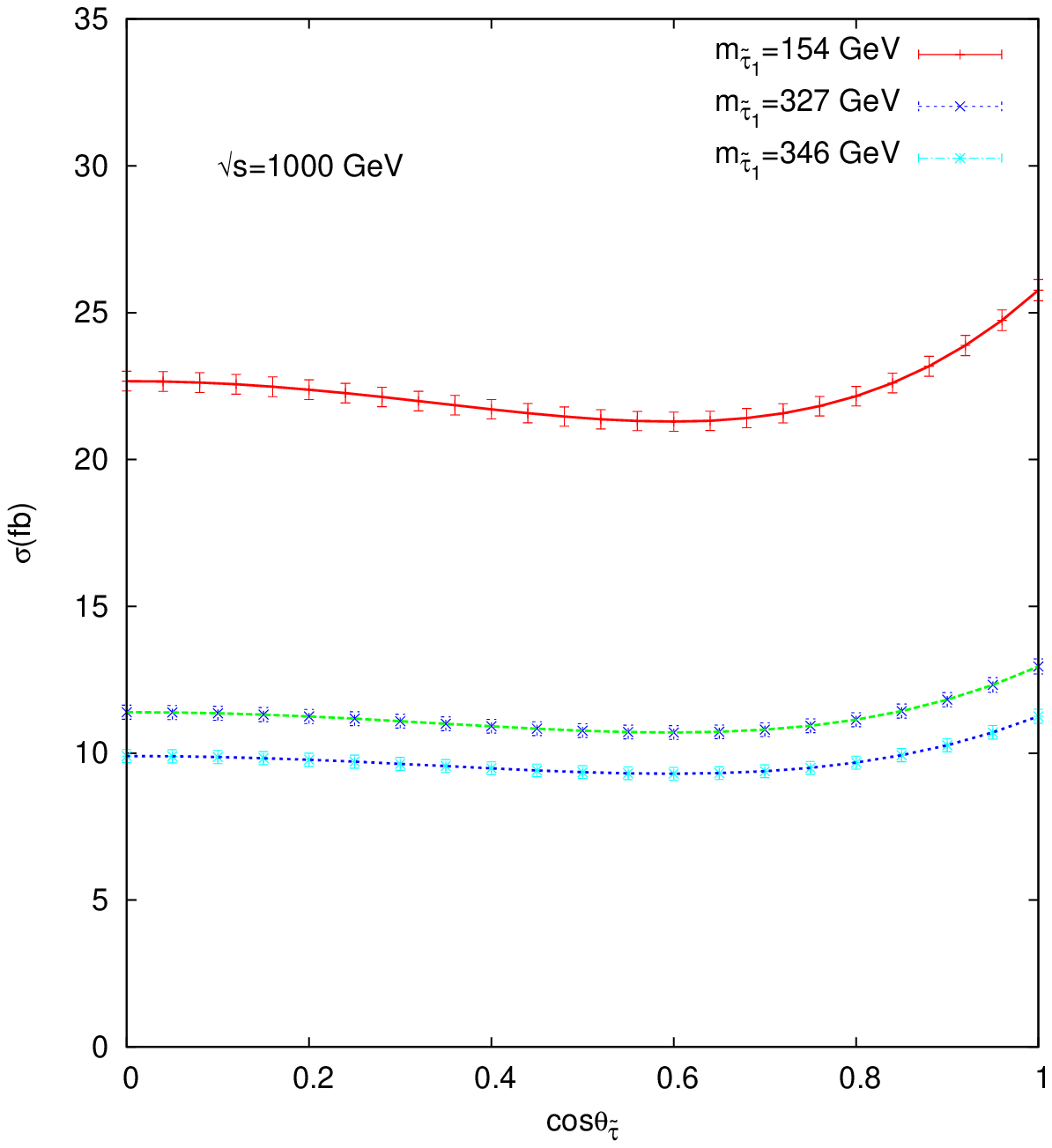}
\includegraphics[
  scale=0.6]{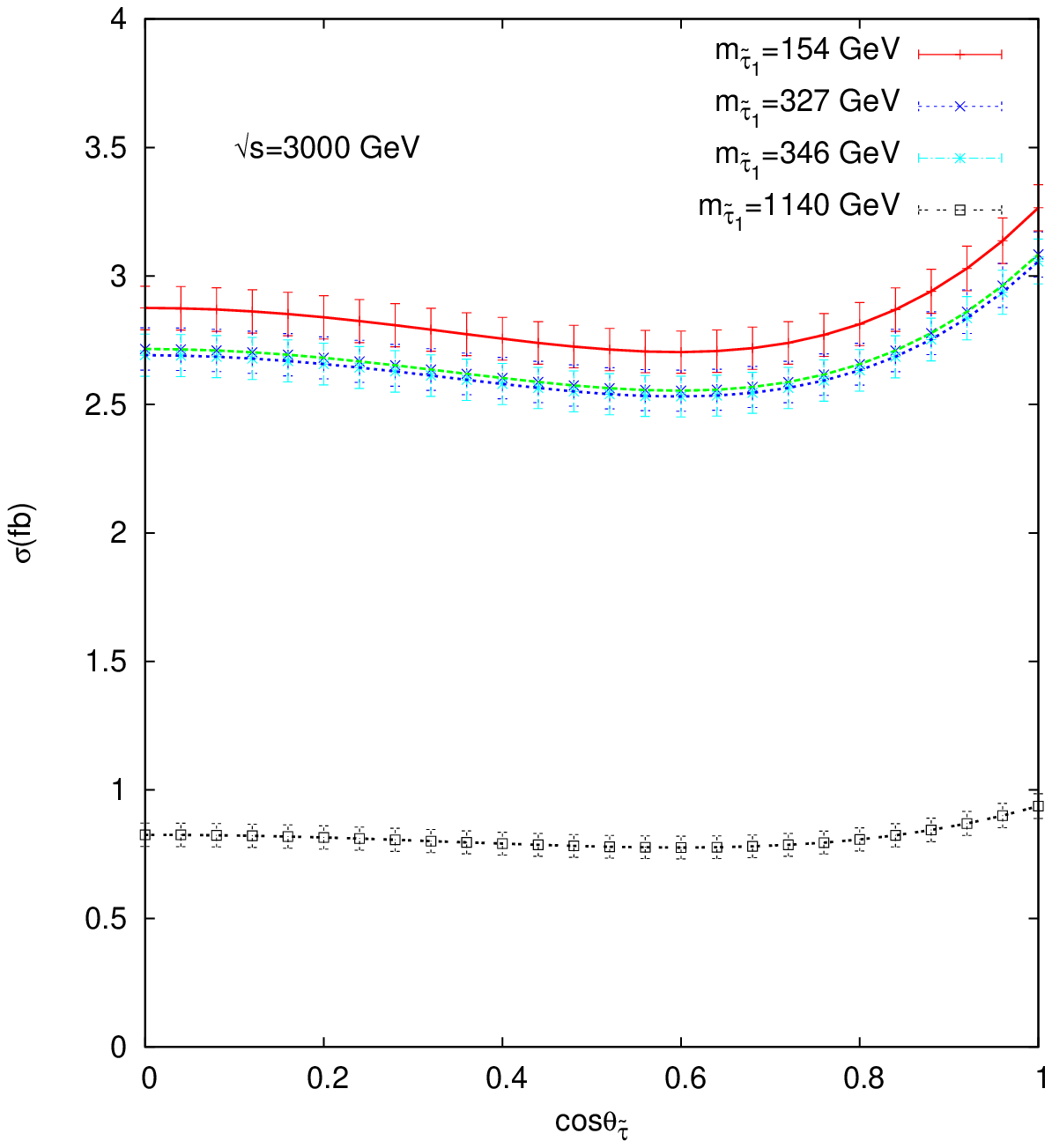}
\includegraphics[
  scale=0.6]{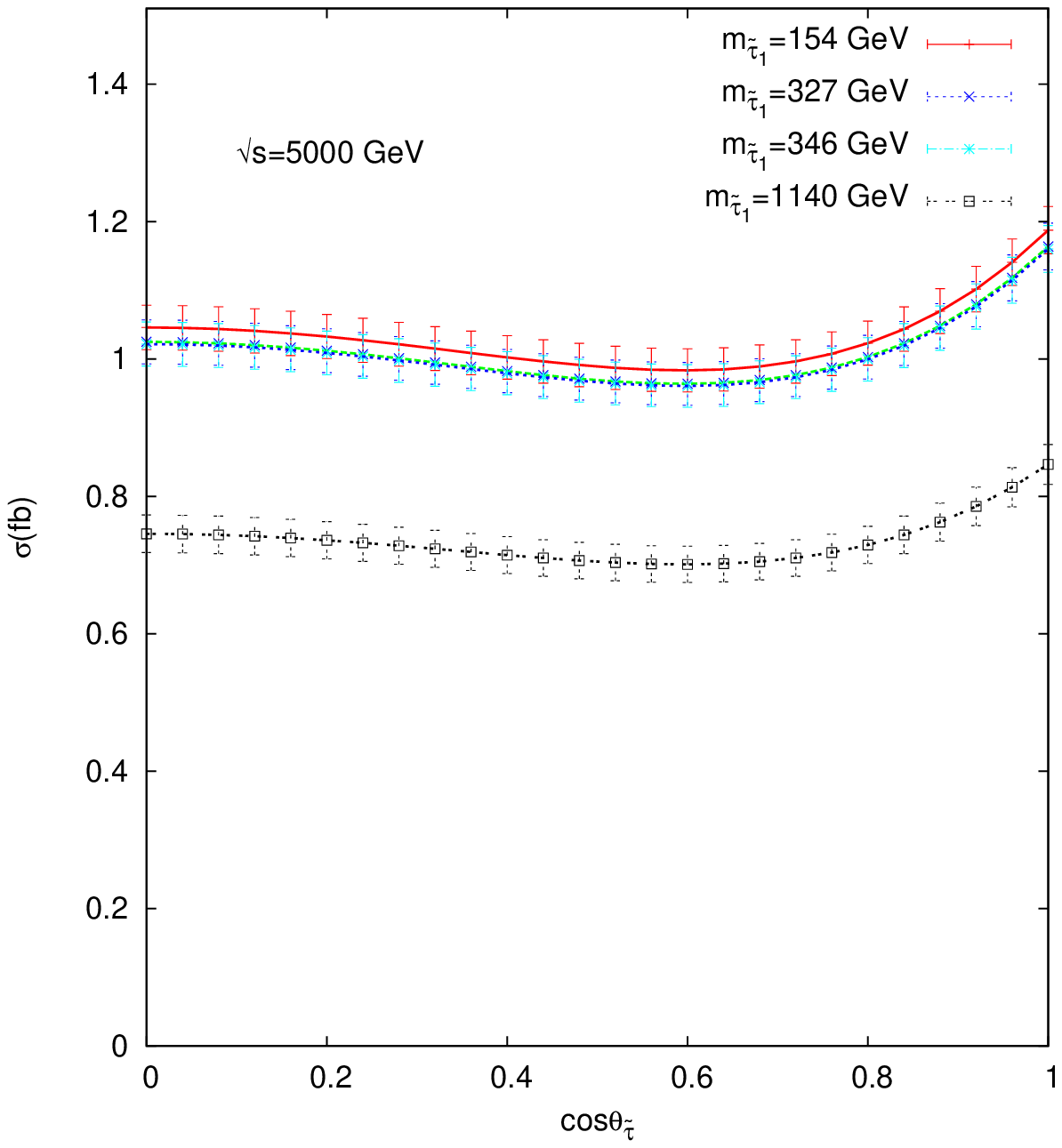}
\caption{\it The unpolarized cross sections for $e^{+}e^{-}\rightarrow\tilde{\tau}_{1}^+\tilde{\tau}_{1}^-$
in fb at $\sqrt{s}=500,1000,3000$ and $5000$ GeV as a function of the stau mixing angle $\theta_{\tilde{\tau}}$ for four of the benchmark scenarios. The expected statistical errors in the cross section measurement are shown for different benchmark points. We take the integrated luminosities $L_{int}=200$ fb$^{-1}$ (for $\sqrt{s}=500$ GeV), $200$ fb$^{-1}$ (for $\sqrt{s}=1000$ GeV), $400$ fb$^{-1}$ (for $\sqrt{s}=3000$ GeV) and $1000$ fb$^{-1}$ (for $\sqrt{s}=5000$ GeV) for the ILC and CLIC energy options.
\label{fig6}}
\end{figure}

The unpolarized cross sections shown in Fig.~\ref{fig6} have
non-trivial dependencies on the stau mixing angle
$\theta_{\tilde{\tau}}$. We see that the cross sections exhibit
minima as functions of $\theta_{\tilde{\tau}}$ when
$\cos\theta_{\tilde{\tau}} \sim 0.6$, due to the interference of
$\gamma$ and $Z$ exchange. One might hope to use this dependence
to extract the stau mixing angle. Unfortunately, the dependencies
of the unpolarized cross sections on $\theta_{\tilde{\tau}}$ are
minimized when $\cos\theta_{\tilde{\tau}} \sim 0$, which
corresponds to a mainly right-handed stau and is the region
expected theoretically, as discussed in the previous section.
Moreover, there is a twofold ambiguity in the determination of
$\theta_{\tilde{\tau}}$, with the alternative value corresponding
to the ${\tilde \tau_1}$ being predominantly left-handed.
Considering the case $m_{\tilde{\tau}_1}=154.0$ GeV, similar to
the mass at the reference point $\epsilon$, and assuming
$\cos\theta_{\tilde{\tau}}=0.60$, for which the cross section
exhibits a minimum and hence no left/right ambiguity, without
radiation effects we find unpolarized cross sections of $48.3\pm
0.5$ pb and $21.3\pm 0.3$ pb for $\sqrt{s}=500$ GeV and
$\sqrt{s}=1000$ GeV with $L_{int}=200$fb$^{-1}$, respectively,
where the errors are statistical.
As seen in Fig.~\ref{fig6}, these
errors on the unpolarized cross-section measurement would not
provide an accurate determination of $\cos\theta_{\tilde{\tau}}$
or of the stau handedness.

\begin{figure}
\includegraphics[scale=0.6]{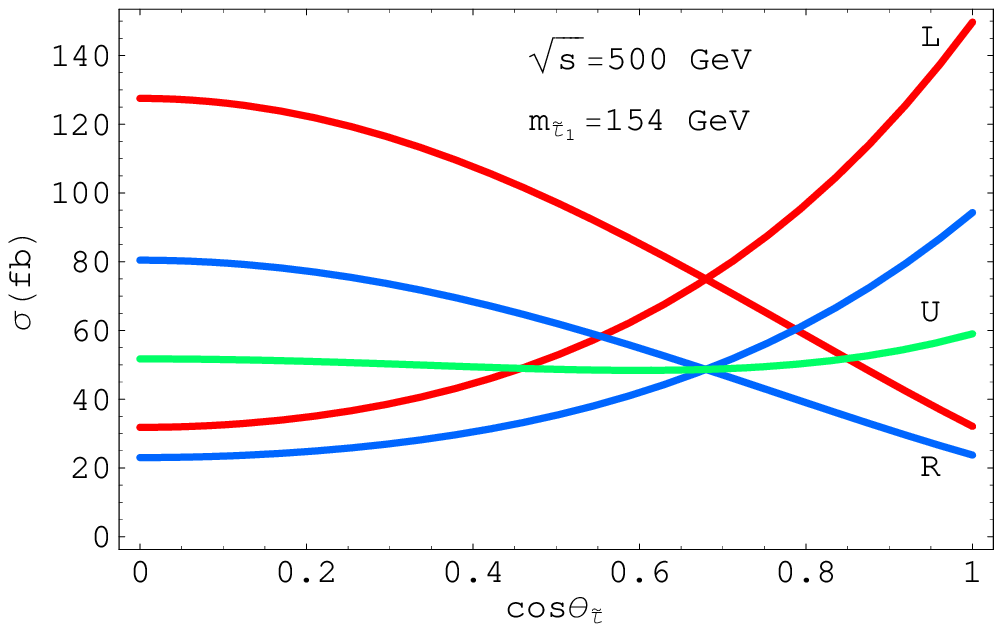}
\includegraphics[scale=0.6]{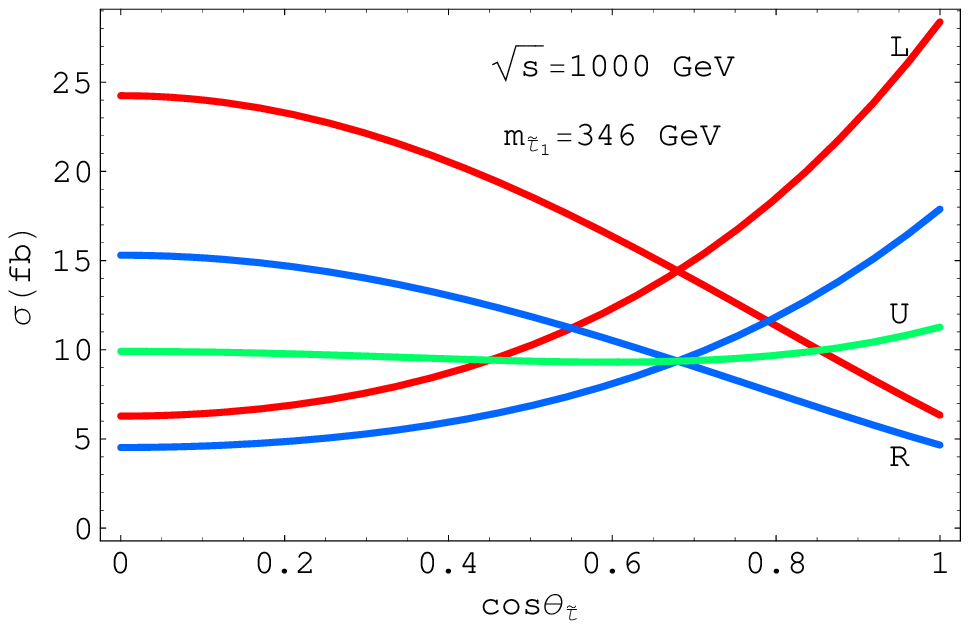}
\includegraphics[scale=0.6]{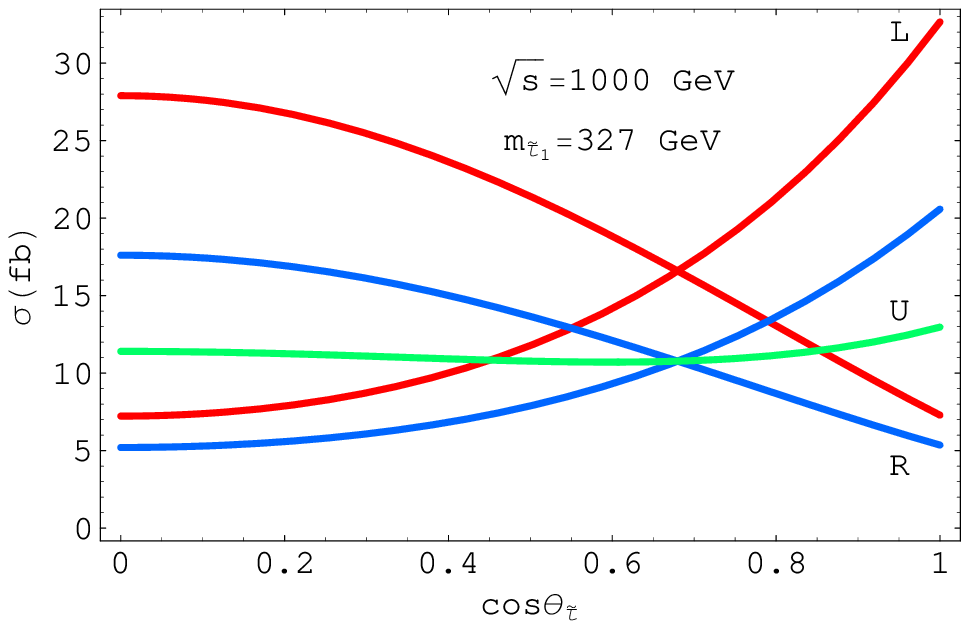}
\includegraphics[scale=0.6]{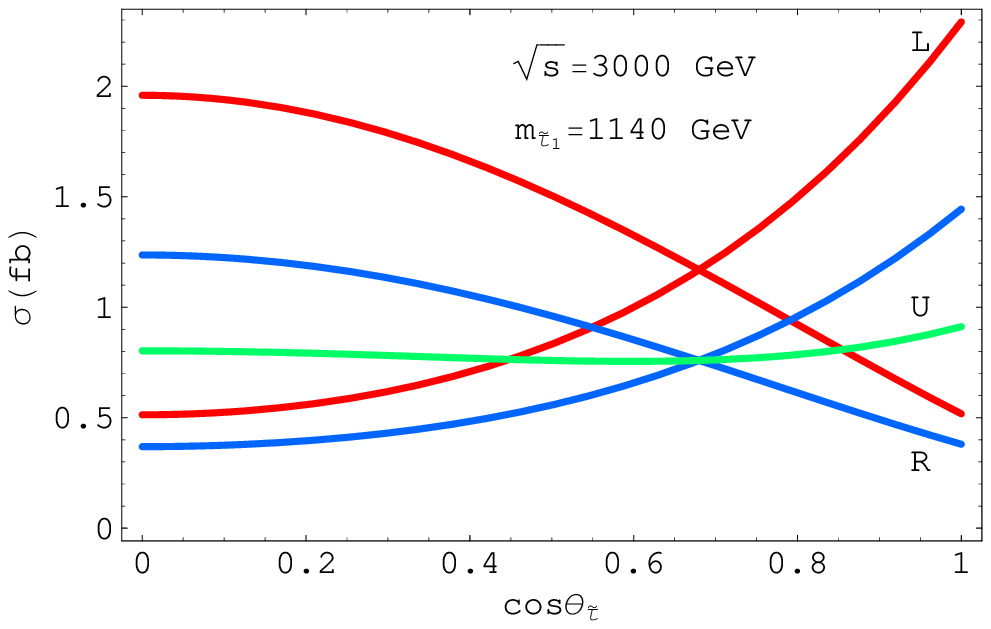}
\caption{\it The cross section of
$e^{+}e^{-}\rightarrow\tilde{\tau}_{1}^+\tilde{\tau}_{1}^-$ in fb
at $\sqrt{s}=500,1000$ and $3000$~GeV as a function of the
stau mixing angle $\theta_{\tilde{\tau}}$, for unpolarized (U) as
well as 90\% left- (L) and 90\% right- (R) polarized $e^-$ beams, and
assuming $m_{\tilde{\tau}_1}=154,346,327$ and $1140$~GeV for the
benchmark points. The red lines show the results for both the electron (90\%) and positron (60\%) polarized,
while blue lines show only electrons (90\%) are polarized.
The statistical experimental errors would be
similar to those shown in Fig.~\ref{fig6}. \label{fig7}}
\end{figure}

One way of determining the handedness of the ${\tilde \tau_1}$ is the $\tau$ decay polarization measurement, as mentioned previously. More detailed analysis of the measurement of tau lepton polarization in hadronic decay channels has been studied in \cite{Nojiri95,Godbole05,Choi07}. The angular distribution of the pion from polarized tau lepton can be expressed as $d\Gamma/\Gamma dz=(1+P_\tau z)/2$, where $P_\tau$ being the polarization of the $\tau$ lepton. The decay $\stau_1\to\tau \tilde{G}$ implies $P_\tau=\sin^2\theta_\tau-\cos^2\theta_\tau$. For the points $\epsilon,\zeta,\eta$ and $\theta$ we have $P_\tau=$$0.902$, $0.965$, $0.959$ and $0.999$, respectively.
The energy distributions of the pions show a maximum in which the minimum of tau energy can be measured.
The fraction of the momenta $p_{\pi^\pm}/p_{\tau-jet}$ can be calculated from
the measurements of pion momentum from tracker and
tau-jet momentum from the calorimetric energy deposition.
Another
would be to use a polarized $e^\pm$ beams.
As we can see from Fig.~\ref{fig7}, the polarization of the $e^\pm$ beam would
offer the possibility of measuring the stau mixing angle (and the stau mass) with much greater
accuracy. Moreover, it can be used to enhance the signal and reduce the backgrounds.
The signal is the same flavor, opposite sign taus with missing transverse momentum $\not{p}_T$.
The SM backgrounds consists of $\tau^+\tau^-+\not{p}_T$ from
the pair production of $W$-bosons and $Z$-bosons as well as the single production $Z+\not{p}_T$.
The total cross section for $W^+W^-$ production is large compared to typical supersymmetry cross sections,
however it proceeds via t-channel exchange of the neutrino, and hence it is very forward peaked.
This background can be much suppressed by using a {\it right-handed} polarized electron beam
and appropriate cuts \cite{Khotilovich05}. Here we note that one can combine
the measurements via threshold scan and polarized cross sections as well
as the tau polarization to complete the determination of the SUGRA model parameters.

\begin{figure}
\includegraphics[
  scale=0.6]{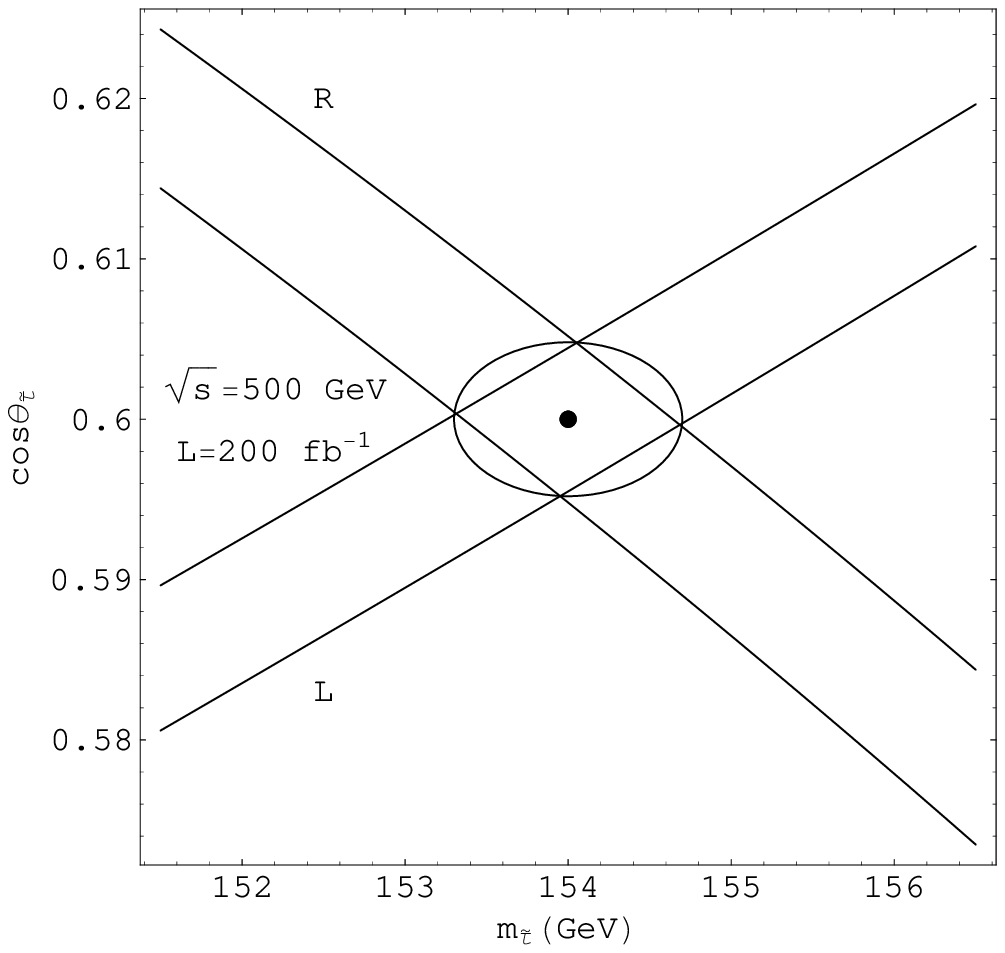}
\includegraphics[
  scale=0.6]{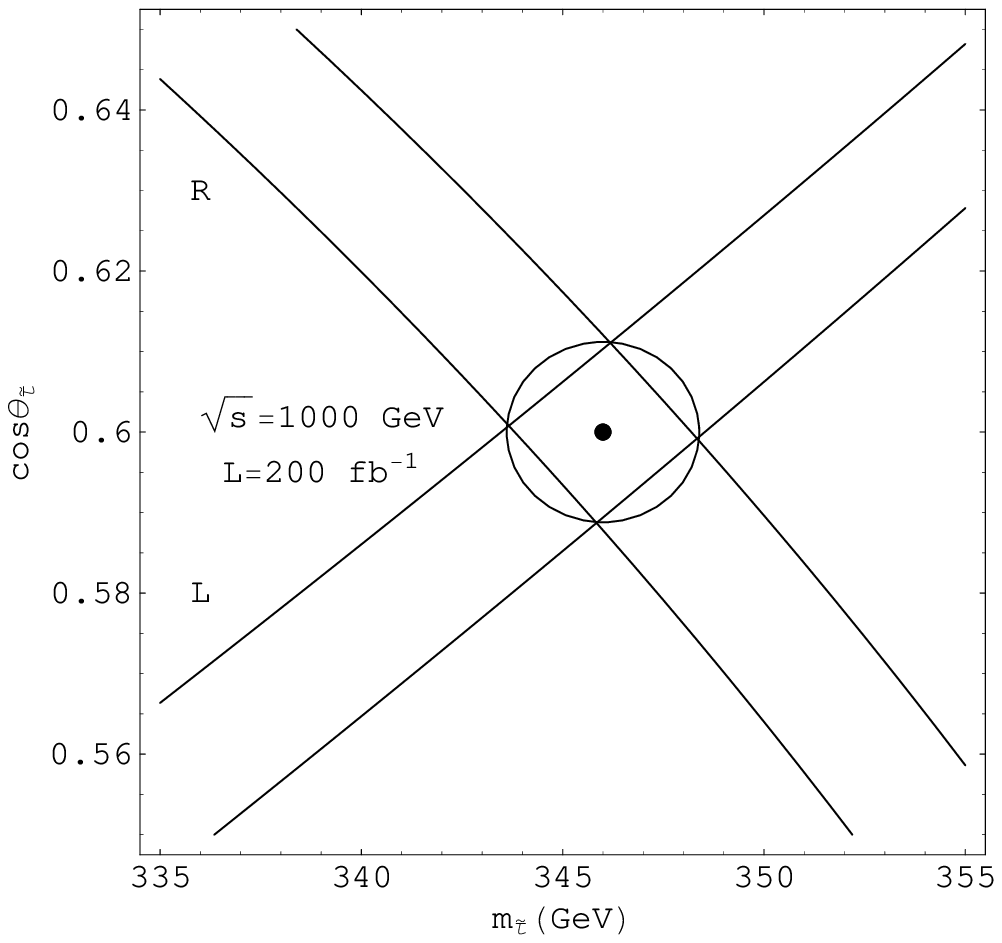}
\includegraphics[
  scale=0.6]{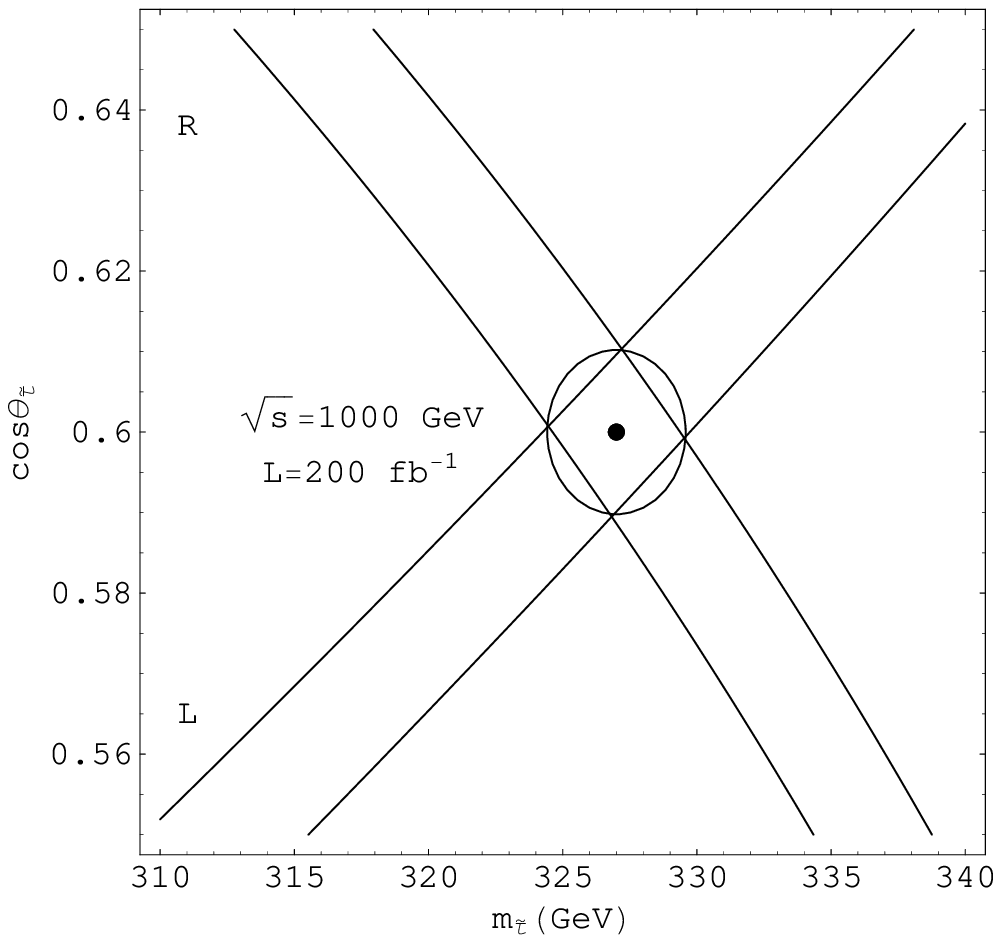}
\includegraphics[
  scale=0.6]{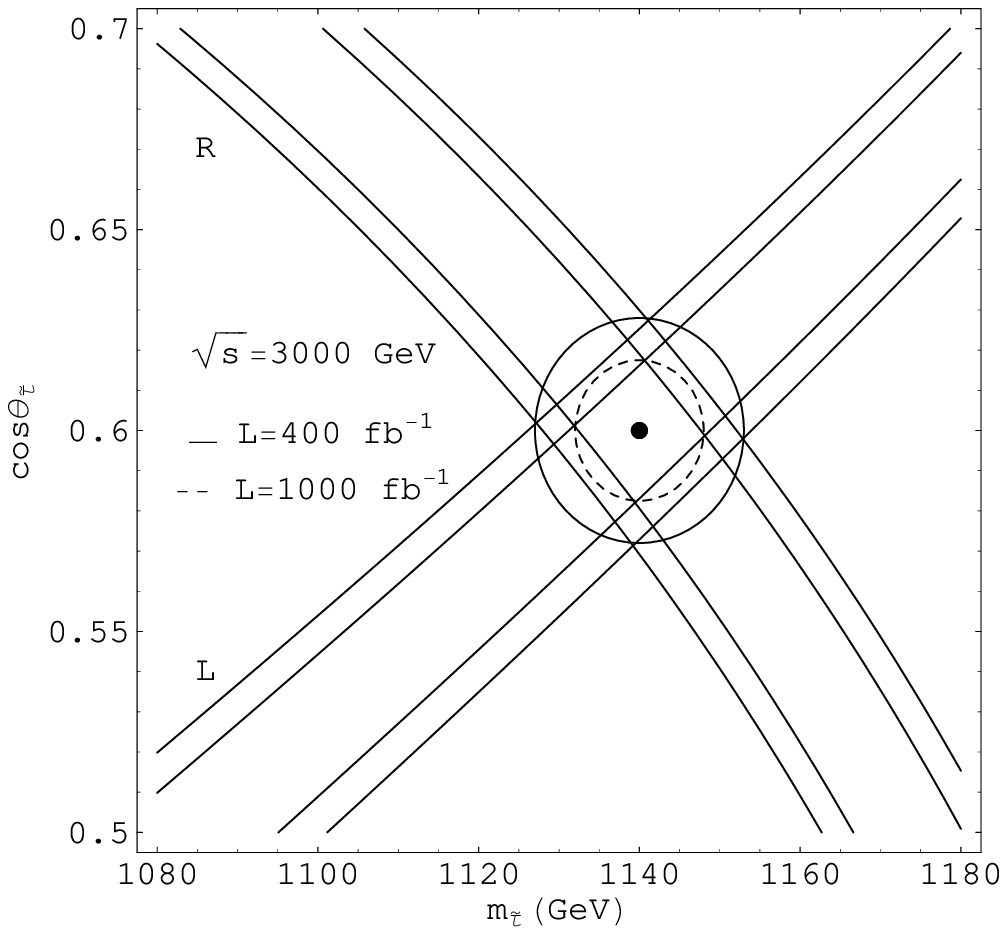}
\caption{\it The error bands for $\cos\theta_{\tilde{\tau}}$ as a
function of the stau mass, as obtained from measurements of the
cross sections for
$e^{+}e^{-}\rightarrow\tilde{\tau}_{1}^+\tilde{\tau}_{1}^-$ at
$\sqrt{s}=500,1000,3000$ GeV with left (90\%)- (L), right (90\%)- (R) and
unpolarized (U) $e^-$ beams, assuming for inputs
$m_{\tilde{\tau}_1}=154,346,327$ and $1140$~GeV, as for the point
$\epsilon,\zeta,\eta,\theta$, and the value of
$\cos\theta_{\tilde{\tau}} = 0.60$. Here we take the positron polarization (60\%)
opposite of electron polarization. \label{fig8}}
\end{figure}

In particular, as seen in Fig.~\ref{fig7}, even a simple measurement of the ratio
of the cross section with left- and right-polarized $e^\pm$ beams would remove the ambiguity
between the left- and right-handed stau solutions for the total unpolarized cross section
measurement that was noted above. Indeed, measurements of the cross sections with
$e^\pm_{L/R}$ beams could determine both $\cos\theta_{\tilde{\tau}}$
and $m_{\tilde{\tau}_1}$ with interesting accuracy.
We display in Fig.~\ref{fig8} the error bands in the $(m_{\tilde{\tau}_1}, \cos\theta_{\tilde{\tau}})$
obtained from measurements of the polarized cross sections
for $e^+e^-\to\tilde{\tau}_1^+\tilde{\tau}_1^-$ at $\sqrt{s}=500,1000,3000$ GeV.
The combined experimental accuracies
in the mass and mixing angle of the stau obtained from this Monte Carlo simulation are
$m_{\tilde{\tau}_1}=154.0\pm 0.7$ GeV and $\Delta\cos{\theta}_{\tilde{\tau}}=0.005$, respectively.
Similar analysis yield $m_{\tilde{\tau}_1}=346.0\pm 2.4$~GeV, $\Delta\cos{\theta}_{\tilde{\tau}}=0.01$ and
$m_{\tilde{\tau}_1}=327.0\pm 2.4$~GeV, $\Delta\cos{\theta}_{\tilde{\tau}}=0.01$ using measurements at $\sqrt{s}=1000$~GeV
and  $m_{\tilde{\tau}_1}=1140\pm 13$~GeV and $\Delta\cos{\theta}_{\tilde{\tau}}=0.03$ using measurements at $\sqrt{s}=3000$~GeV,
for the points $\zeta$, $\eta$ and $\theta$, respectively. Here, we take the integrated luminosities $L_{int}=200$ fb$^{-1}$ for $\sqrt{s}=500$(or $1000$) GeV and $L_{int}=400$ fb$^{-1}$ for $\sqrt{s}=3000$ GeV. For a high luminosity option ($L_{int}=1000$ fb$^{-1}$) for the CLIC with $\sqrt{s}=3000$ GeV, the stau $\tilde{\tau}_1$ can be measured with less errors on the mass and mixing, namely $\Delta m_{\tilde{\tau}}=8$ GeV and $\Delta\cos{\theta}_{\tilde{\tau}}=0.02$ at the point $\theta$. In our calculations, we take into account the degree of polarization $P_{-}=90\%$ for the electron (which means 90\% of the electrons
are left-polarized and the rest is unpolarized) and $P_+=60\%$ for the positron. It might be possible to reduce
further the error in $m_{\tilde{\tau}_1}$ by using tracking and time-of-flight
measurements, as has been considered for the LHC, but we see from Fig.~\ref{fig8}
that this would be unlikely to improve significantly the determination of $\cos\theta_{\tilde{\tau}}$.

\section{Detection of Stau Decays}

Measuring stau decays using staus that stop and decay inside a linear-collider
detector was first discussed in~\cite{Martyn06}. Measurement of, e.g., the mass
of the gravitino and the decay tau polarization
require optimizing the number of staus stopped and then decaying inside the
collider detector, which we discuss here. As in~\cite{Martyn06},
we assume a future linear-collider detector with the
structure proposed in~\cite{TESLA01}. A metastable stau may stop in the hadron
calorimeter or in the iron yoke. The amount of absorber
material $R$(g/cm$^2$) in the detector, summed along the radial direction at
different longitudinal angles, is 835 to 1250~g/cm$^{2}$ for
the hadronic calorimeter (HCAL) and 1810 to 2750~g/cm$^{2}$ for the magnet
return yoke, respectively. As discussed in~\cite{Martyn06}, the range $R$
of a metastable stau with given $\beta\gamma$ and mass $m_{\tilde{\tau_1}}$ is given by
\begin{equation}
log_{10}(\beta\gamma) \; = \; \frac{\left[ log_{10}(R/m_{\tilde{\tau_1}}) - c_{1} \right]}{c_{2}},
\label{range}
\end{equation}
where $c_{1}=2.087$ and $c_{2}=3.227$ for steel~\cite{Rossi52}.
The corresponding values of $\beta\gamma$ for staus stopping
in different detector parts for the benchmark points are shown in Table~\ref{table4}.
(The corresponding values for some other stau masses are given in Table~II of~\cite{Martyn06}.)
As a representative example, we consider below the production of slow-moving staus
with $\beta \gamma < 0.4$, that might plausibly be stopped with a linear-collider
detector, so that their decays could be observed.

\begin{table}
\caption{\it The maximal values of $\beta\gamma$ below which staus
would be stopped in different detector parts, for the benchmark scenarios $\epsilon, \eta, \zeta$
and $\theta$.
\label{table4}}
\begin{tabular}{|c|c|c|}
\hline
$m_{\tilde{\tau}}$ (GeV)&
$\beta\gamma$ (HCAL)&
$\beta\gamma$ (Iron Yoke)\tabularnewline
\hline
154&
0.38-0.43&
0.48-0.55 \tabularnewline
\hline
346&
0.30-0.34&
0.38-0.43 \tabularnewline
\hline
327&
0.30-0.34&
0.38-0.44  \tabularnewline
\hline
1140&
0.21-0.23&
0.26-0.30 \tabularnewline
\hline
\end{tabular}
\end{table}

\begin{figure}
\includegraphics[width=0.45\textwidth]{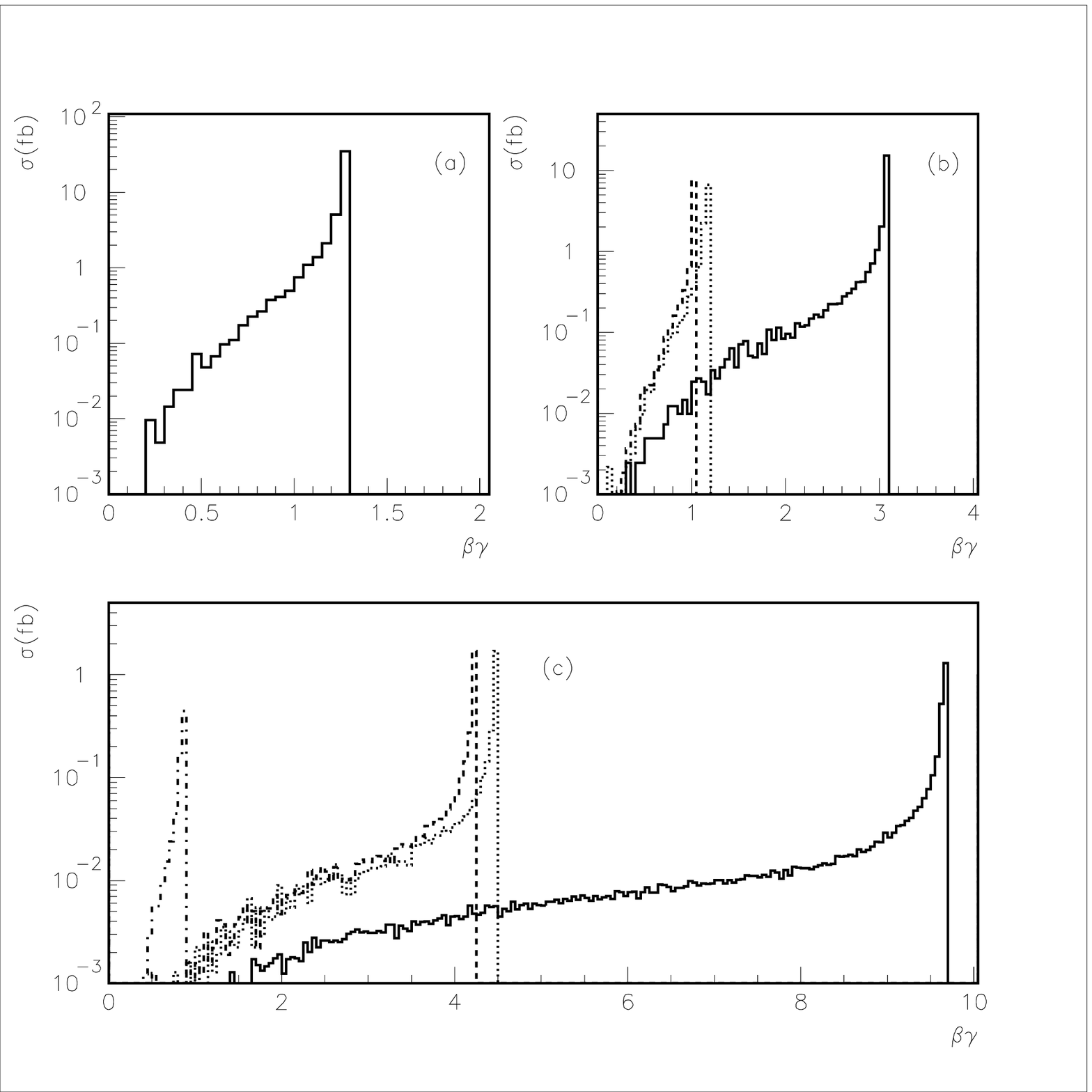}
\includegraphics[width=0.45\textwidth]{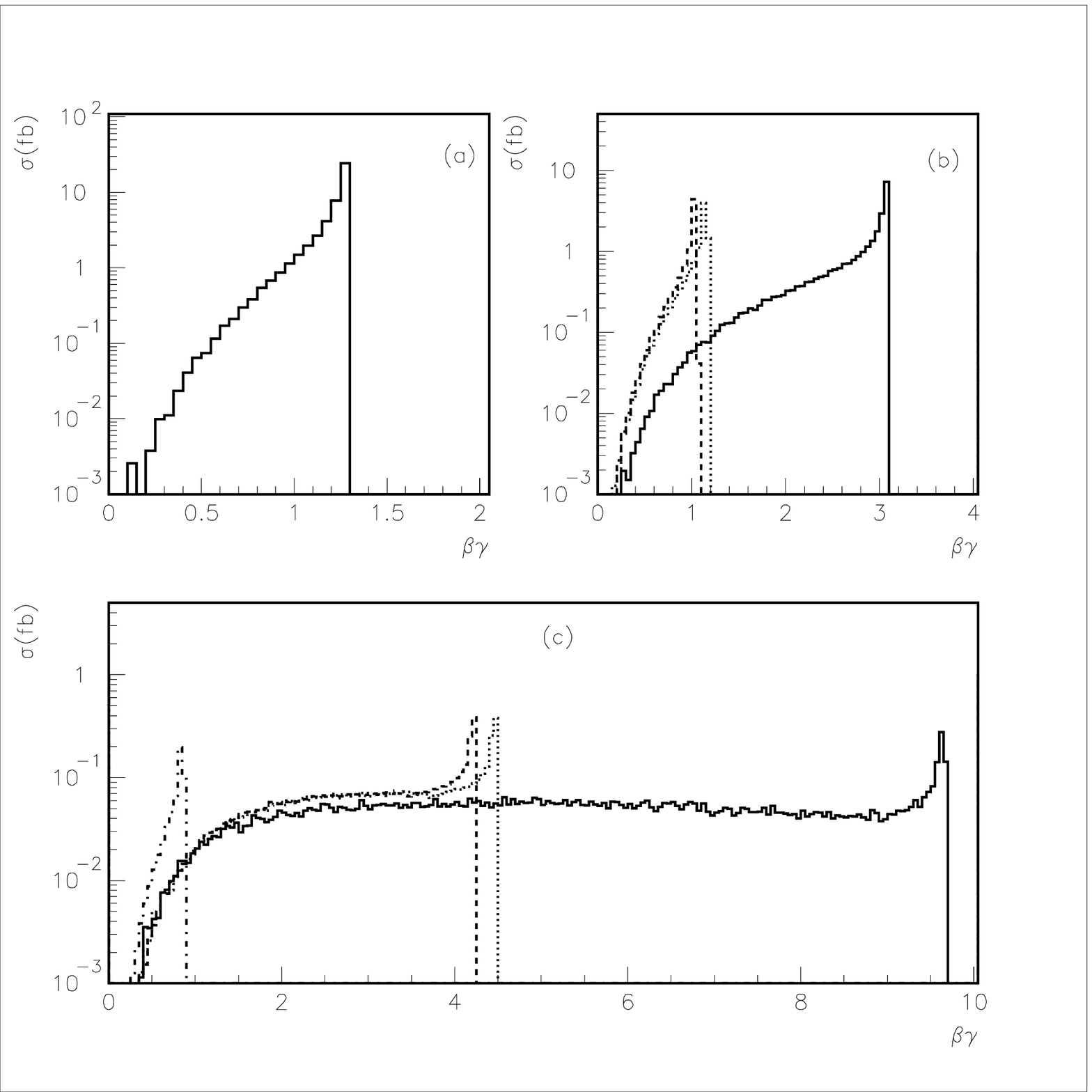}
\caption{\it The $\beta\gamma$ distributions for direct stau
pair-production
$e^{+}e^{-}\rightarrow\tilde{\tau}_{1}^+\tilde{\tau}_{1}^-$ at
different center-of-mass energies: a) $\sqrt{s}=500$ GeV, b)
$\sqrt{s}=1000$ GeV and c) $\sqrt{s}=3000$ GeV. The solid, dashed,
dotted and dashed-dotted lines correspond to the gravitino DM
points $\epsilon$, $\zeta$, $\eta$ and $\theta$, respectively.
These distributions were obtained from simulations of running
conditions designed (left) to optimize the luminosity close to the
nominal center-of-mass energy and (right) to optimize the total
luminosity. The low-energy tails of the $\beta\gamma$
distributions are apparent, and much larger in the right-hand
panel. Here, we have used the {\tt CALYPSO}
library~\cite{Daniel01} to include the luminosity spectrum into
the event generator {\tt PYTHIA}. \label{fig9}}
\end{figure}

In the absence of photon radiation, the velocity of the outgoing staus in
the direct pair-production reaction $e^{+}e^{-}\rightarrow\tilde{\tau}_{1}^+\tilde{\tau}_{1}^-$
is simply $\beta =  \sqrt{1-4m_{\tilde{\tau}}^{2}/s}$. However, the values of the stau velocities
are altered when initial-state radiation and a realistic luminosity spectrum are
taken into account. Fig.~\ref{fig9} shows
the distributions of $\beta\gamma$ expected for the $\tilde{\tau}_{1}$s
pair-produced in the various gravitino DM scenarios at the center-of-mass
energies 500 GeV (the design for the ILC), 1 TeV and 3 TeV (the design for CLIC).
The two sets of $\beta\gamma$ distributions correspond to alternative luminosity spectra
yielded by beam conditions designed to optimize the luminosity close to
the nominal center-of-mass energy (left) and the total luminosity (right).
We see that the $\beta\gamma$ distributions exhibit substantial
low-energy tails, particularly in the case where the total luminosity is optimized, as
shown in the right panel of Fig.~\ref{fig9}.

We plot in Fig.~\ref{fig5} the cross sections the production of staus with
$\beta\gamma < 0.4$ via direct pair-production $e^{+}e^{-}\rightarrow\tilde{\tau}_{1}^+\tilde{\tau}_{1}^-$
accompanied by initial- and final-state radiation, which opens up the possibility of producing
slow-moving staus. We see that the optimal center-of-mass energies for
producing slow-moving staus via direct pair-production are
330, 730, 700 and 2500~GeV for benchmarks $\epsilon, \zeta, \eta$
and $\theta$, respectively. We note that the center-of-mass energy dependencies of these
cross sections for slow-moving staus are completely different from the total cross sections
shown in Fig.~\ref{fig5}, reflecting the increasing difficulty of radiating sufficiently to produce a
slow-moving stau as the center-of-mass energy increases.

Table~\ref{table5} gives estimates of the numbers of staus with
$\beta \gamma < 0.4$, as produced directly via
$e^{+}e^{-}\rightarrow\tilde{\tau}_{1}^+\tilde{\tau}_{1}^-$
accompanied by initial- and final-state radiation at selected
nominal center-of-mass energies. These staus might stop in a
typical linear-collider detector, where their decays could then be
observed. We note that the numbers of stopped staus obtained at
the optimal center-of-mass energies for benchmark points are more
than an order of magnitude larger than the numbers that would be
stopped in measurements at center-of-mass energies corresponding
(approximately) to the peaks of the total pair-production cross
sections, namely 500, 1000, 1000 and 3000~GeV, respectively.

\begin{table}
\caption{\it The numbers of staus with $\beta \gamma < 0.4$, that may be
stopped in a generic linear-collider detector, for the different benchmark
points at different center-of-mass energies and luminosities. In the fourth column, the energies are
optimized for the direct pair-production of slow-moving staus, whereas in the last column
all sparticle pair-production cross sections are included.
\label{table5}}

\begin{tabular}{|c|c|c|c|c|c|}
\hline $\sqrt{s}$(GeV)& $500$& $1000$& $3000$& Optimal for& Maximal including
\tabularnewline $L_{int}$(fb$^{-1}$)& 200&200 &400(1000) & pair prod'n & other prod'n processes
\tabularnewline \hline $\epsilon$& 34& 4& 4(10) &1500&
1700 \tabularnewline \hline $\zeta$& -& 12& 4(10) & 254 & 700
\tabularnewline \hline $\eta$& -& 10& 4(10) &370 & 600 \tabularnewline
\hline $\theta$& -& -& 8(20) &56(140) & 140(350) \tabularnewline \hline
\end{tabular}
\end{table}

\section{Indirect Stau Production}

\begin{figure}[t]
\includegraphics[width=8cm]{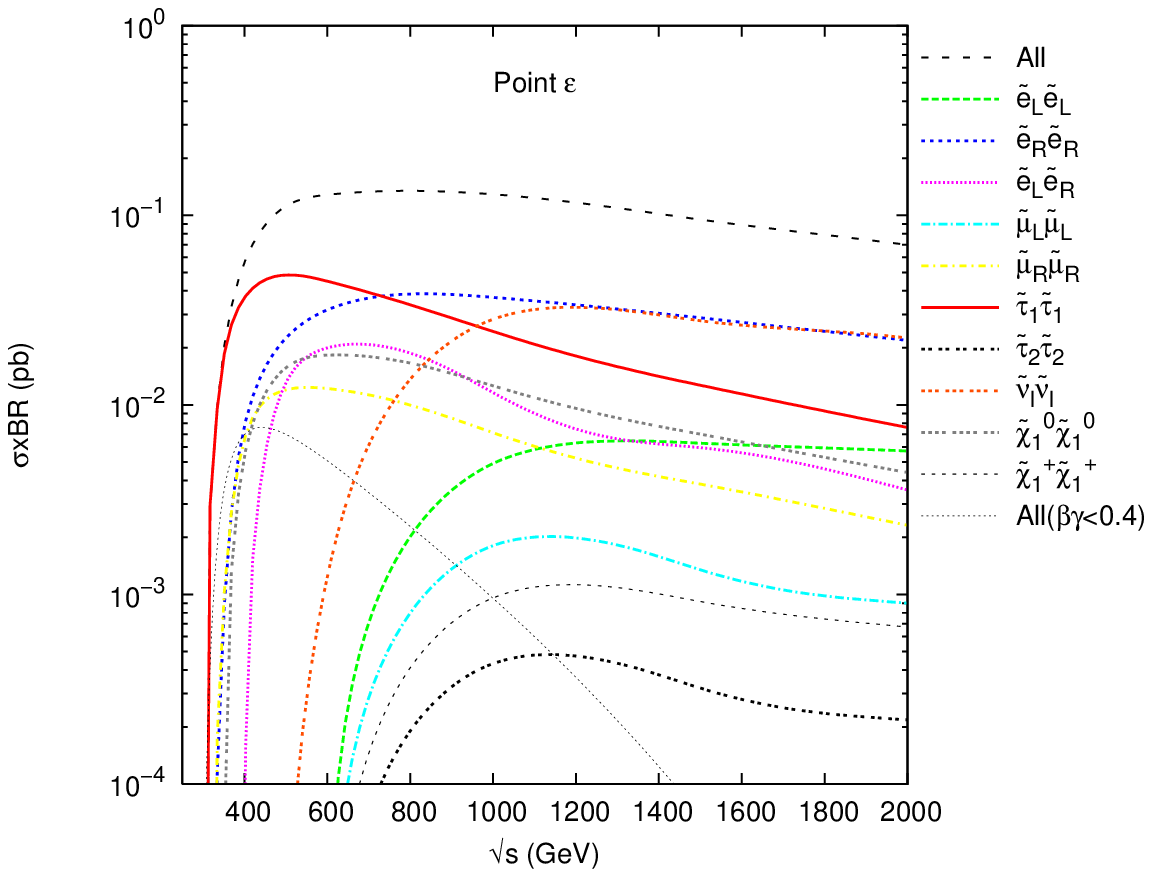}
\includegraphics[width=8cm]{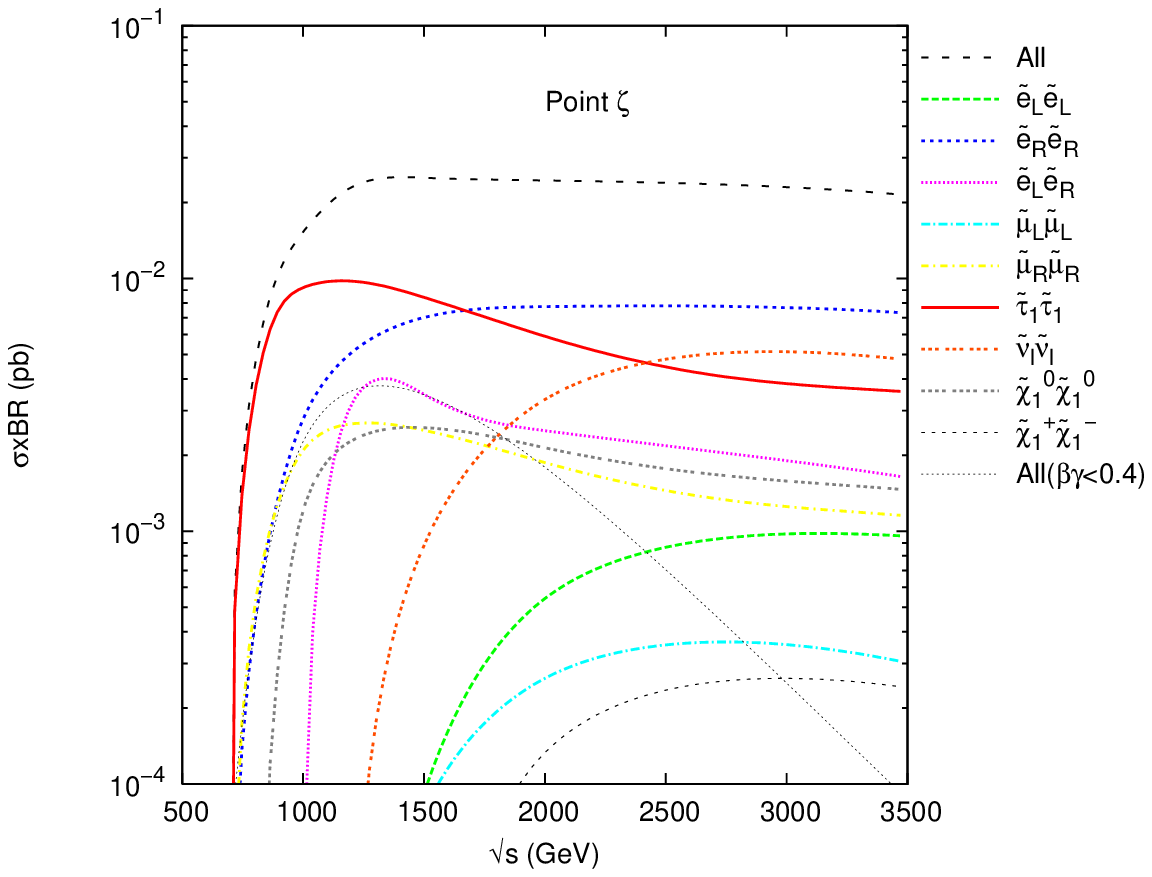}
\includegraphics[width=8cm]{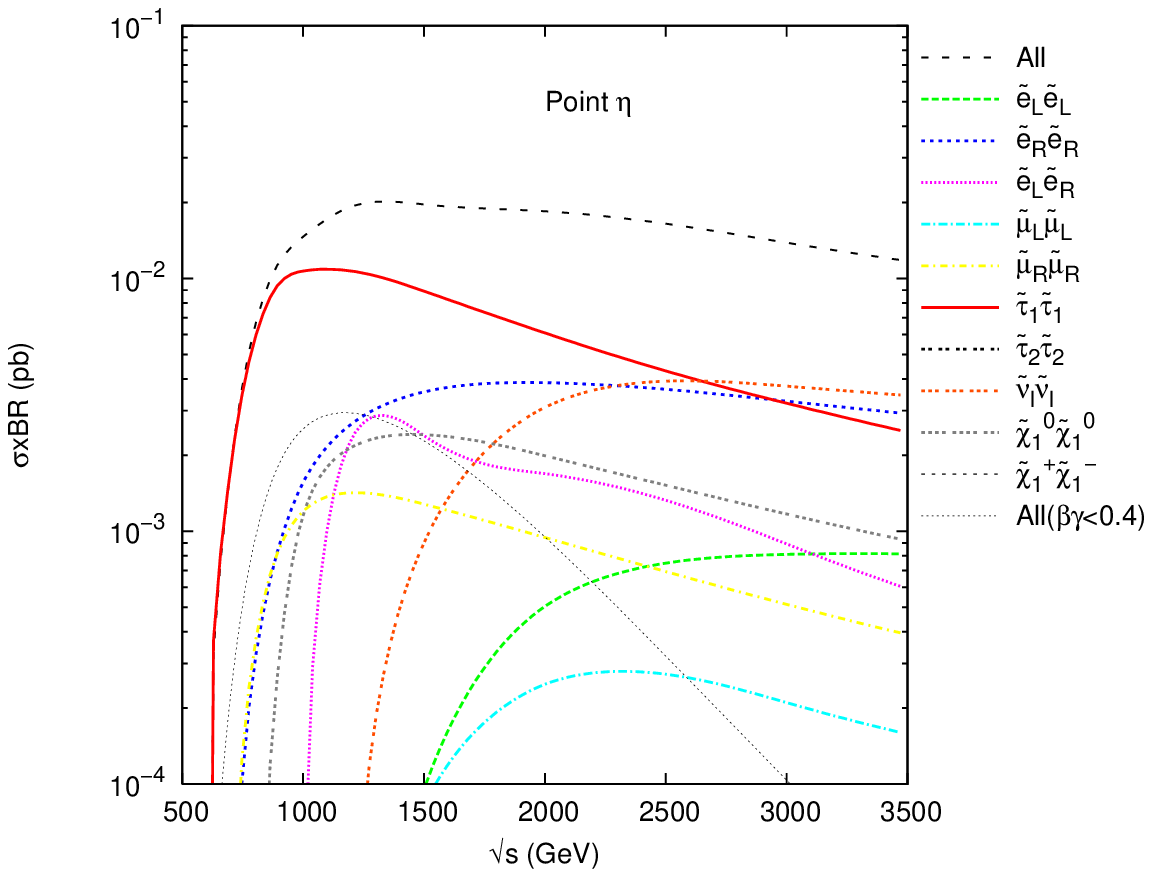}
\includegraphics[width=8cm]{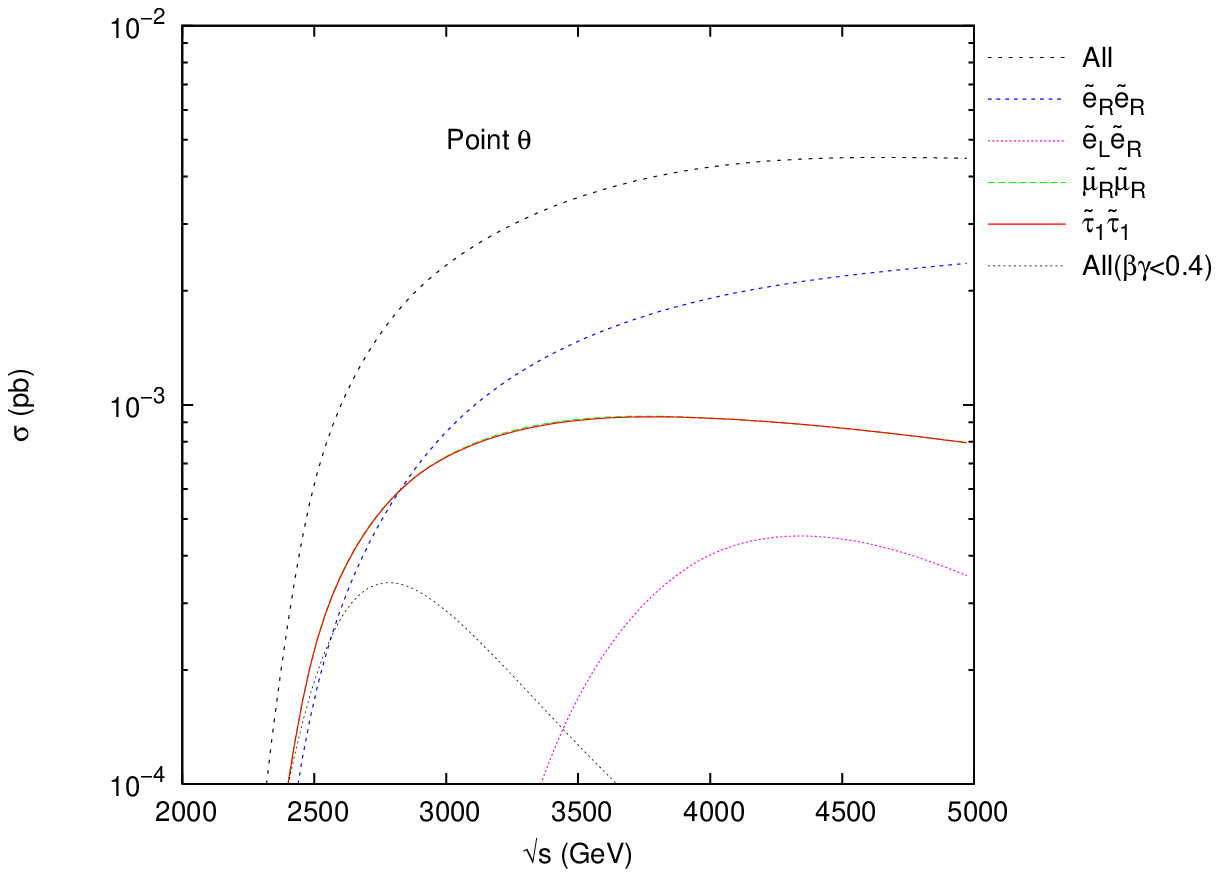}
\caption{\it The cross sections for pair-production of supersymmetric particles in the
benchmark scenarios $\epsilon, \zeta, \eta$ and $\theta$, as
functions of the $e^+ e^-$ center-of-mass energy. Also shown are the
yields of slow-moving staus with $\beta \gamma < 0.4$.
\label{fig10}}
\end{figure}

In addition to direct stau pair-production
$e^{+}e^{-}\rightarrow\tilde{\tau}_{1}^+\tilde{\tau}_{1}^-$,
staus can be produced via the cascade decays
of heavier sparticles, e.g. sleptons, $\tilde{e}_{R}\to \tilde{\tau}_{1}\tau e$,
or neutralinos, $\tilde{\chi}_{1}^{0}\to\tilde{\tau}_{1}\tau$.
Each heavier sparticle eventually yields one stau among its cascade
decay products, which later decays into a gravitino.
For example, for the nominal ILC center-of-mass energy of $\sqrt{s}=500$ GeV,
the most important contributions to stau-pair production at point $\epsilon$
in fact come from the processes $e^+e^-\to\tilde{\chi}_{1}^{0}\tilde{\chi}_{1}^{0}$
and $e^+e^-\to\tilde{e}_R^{+}\tilde{e}_{R/L}^{-}$
with cross sections of $1.26\times 10^{-1}$~pb and
$7.44\times 10^{-2}/6.76\times 10^{-2}$ pb, respectively.
On the other hand, for the point $\theta$ at $\sqrt{s}=3000$ GeV,
the main contribution comes from the pair production of
right-handed selectrons, smuons and neutralinos with the cross
sections $8.49\times 10^{-4}$ pb, $7.31\times 10^{-4}$ pb and
$9.66\times 10^{-4}$ pb, respectively. We display in
Fig.~\ref{fig10} the dominant cross sections for sparticle
pair-production in unpolarized $e^{+}e^{-}$ collisions in
benchmarks $\epsilon, \eta, \zeta$ and $\theta$, as functions of
the center-of-mass energy.

The pair-production of these heavier sparticles would be interesting in its own right,
providing new opportunities to study sparticle spectroscopy with high precision, and
to measure many sparticle decay modes. Some prominent decay modes of
selectrons, smuons, neutralinos, charginos and heavier staus are shown in
Table~\ref{table6}~$^3$\footnotetext[3]{In the case of point $\theta$, the small
${\tilde e}_R - {\tilde \tau}_1$
mass difference implies that the only three-body decays allowed kinematically are
${\tilde e}_R \to {\tilde \tau}_1 {\bar \nu} \nu$, and similarly for ${\tilde \mu}_R$.}:
measuring these would give much information about sparticle masses and couplings, and hence
the underlying pattern of supersymmetry breaking.

\begin{table}
\caption{\it Significant decay modes and branching ratios of selectrons,
smuons, neutralinos, charginos and heavy staus for the chosen
benchmark points $\epsilon$,$\zeta$,$\eta$ and $\theta$. Here, the charged lepton $l^-$ may be either $e^-$ or $\mu^-$, and the branching ratios add up to $\sim 100$ \% when these and
charge-conjugate modes are added.}
\label{table6}
\begin{tabular}{|c|c|c|}
\hline
Sparticle&
Significant decay modes&
BR (\%) for $\epsilon$,$\zeta$,$\eta$,$\theta$\tabularnewline
\hline
$\tilde{l}_{L}^{-}/\tilde{\nu}_{l}$&
$\tilde{\chi}_{1}^{0}l^{-}/\tilde{\chi}_{1}^{0}\nu_{l}$&
$100,100,100,100$\tabularnewline
\hline
$\tilde{l}_{R}^{-}$&
$\tilde{\tau}_{1}^{-}\tau^{+}l^{-}$&
$45,42,40,0$\tabularnewline
&
$\tilde{\tau}_{1}^{+}\tau^{-}l^{-}$&
$55,58,60,0$\tabularnewline
&
$\tilde{\tau}_{1}^{-}\bar{\nu}_\tau\nu_{l}$&
$0,0,0,100$\tabularnewline
\hline
$\tilde{\nu}_{\tau}$ & $\tilde{\chi}_{1}^{0}\nu_\tau$&
$55,21,17,64$\tabularnewline
& $\tilde{\tau}_1^-W^+$ & $45,79,83,36$\tabularnewline
\hline
$\tilde{\tau}_{2}^{-}$&
$\tilde{\chi}_{1}^{0}\tau^{-}$&
$50,20,17,56$\tabularnewline
& $\tilde{\tau}_1^-Z^0$ & $23,37,39,16$\tabularnewline
& $\tilde{\tau}_1^-H^0$ & $27,43,44,28$\tabularnewline
\hline
$\tilde{\chi}_{1}^{0}$&
$\tilde{\tau}_{1}^{\pm}\tau^{\mp}$&
$35,30,29,17$\tabularnewline
&
$\tilde{l}_{R}^{\pm}l^{\mp}$&
$7,10,11,17$\tabularnewline
\hline
$\tilde{\chi}_{1}^{-}$&
$\tilde{\tau}_{1}^{-}\nu_{\tau}$&
$13,4,5,0$\tabularnewline
&$\tilde{\nu}_{l}l^{-}$&
$18,16,16,17$\tabularnewline
&$\tilde{l}^{-}\nu_{l}$&
$9,13,14,16$\tabularnewline
&$\tilde{\nu}_{\tau}\tau^-$&
$22,19,19,18$\tabularnewline
&$\tilde{\tau}_{2}^-\nu_{\tau}$&
$8,15,15,17$\tabularnewline
\hline
$\tilde{\chi}_{2}^{0}$&
$\tilde{\tau}_{1}^{\pm}\tau^{\mp}$&
$7,2,2,0$\tabularnewline
&
$\tilde{\nu}_{\tau}\bar{\nu}_\tau$/$\tilde{\bar{\nu}}_{\tau}\nu_\tau$&
$10,10,10,9$\tabularnewline
&
$\tilde{\nu}_{l}\bar{\nu}_l$/$\tilde{\bar{\nu}}_l\nu_l$&
$9,8,8,8$\tabularnewline
&
$\tilde{l}^\pm l^\mp$&
$5,7,7,8$\tabularnewline
&
$\tilde{\tau}_{2}^{\pm}\tau^\mp$&
$4,8,8,8$\tabularnewline
\hline
\end{tabular}
\end{table}

\begin{figure}[t]
\includegraphics[width=10cm,height=8cm]{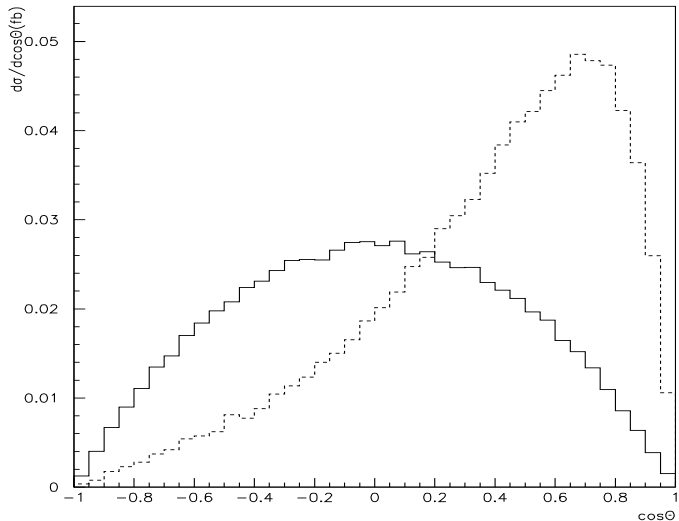}
\caption{\it The angular distributions of the
staus (solid line) and right-handed selectrons (dashed line)
from their pair productions at
the center of mass energy $\sqrt{s}=3000$ GeV.
\label{fig11}}
\end{figure}

Concerning the point $\theta$,
the angular distributions of the staus and right-handed selectrons
produced directly are given in Fig. \ref{fig11}. The staus
peak around the central part of the distribution for
their pair production while selectrons place forward direction for
the $\tilde{e}^+_R\tilde{e}^-_R$ production due to the
contribution from the neutralino exchange in the $t-$channel.
If one study only the direct production of staus it is useful
to apply a cut on the angular
distribution $\cos\theta<2/3$ for an easy identification of staus.
In this case the direct signal
cross section reduces $15\%$ and the contribution from
indirect production can be eliminated mostly.
The right-handed smuons show the same angular distribution
as the staus for their pair production.
At this point, in order to be sure that stau is still
an NLSP we need to have experimental identification of
the corresponding final states.
Assuming the stau is an NLSP we expect hadronic
decay channels of tau leptons and the missing energy,
otherwise smuons give their unique signatures in the muon
spectrometer when they stopped inside the detector.
However, here we take the stau as an NLSP with a good
motivation from the BBN implications.
We show in Fig. $\ref{fig10}$ all the
contributions to the stau pairs
from the indirect production
mechanisms for the corresponding benchmark points.
In addition, our interest here is in the contributions
of these heavier sparticles to the total yields of slow-moving
staus with $\beta \gamma < 0.4$, which are also shown in Fig.~\ref{fig10}.
Comparing these yields with those produced by direct
pair-production alone, shown in Fig.~\ref{fig5},
we see that they are greatly enhanced.
It is also apparent
that the optimal center-of-mass energies are
substantially higher, reflecting the higher
thresholds for pair-producing heavier sparticles.
Specifically, we find optimal
center-of-mass energies of 430, 1200, 1150 and 2800~GeV
for the benchmark points $\epsilon, \zeta, \eta$
and $\theta$, respectively. The last column of
Table~\ref{table5} displays the maximal number of
slow-moving and hence stoppable
staus that might be obtainable at these optimal
center-of-mass energies in the four
benchmark scenarios. These numbers provide greatly
improved prospects for
observing stau decays into gravitinos. This would yield
direct insight into the
mechanism of supersymmetry breaking,
via a measurement of the stau lifetime, and
possibly of the gravitino mass,
via the kinematics of stau decays.

\section{Conclusion}

We have discussed in this paper how the study of a metastable stau
NLSP could be optimized at a linear collider such as the ILC or CLIC.
The stau production rate, particularly using a polarized $e^-$ beam,
would provide information on the handedness of the stau, as
well as its mass. The center-of-mass energy could then be optimized for the
combined direct and indirect production of
slow-moving staus that would stop in the detector. Their decays
could provide valuable information, not only on the stau lifetime and on the
mass of the gravitino, but also supplementary information
on the handedness of the stau via a
measurement of the polarization of the tau decay products.
The following are possible scenarios for studying metastable stau
scenarios at the LHC and a subsequent linear collider.

If there is a metastable stau weighing
up to $\sim 300$~GeV, as in benchmarks $\eta$ and $\zeta$, the LHC has
been shown able to produce and detect them, and also
to analyze the mSUGRA models in which they appear~\cite{Ellis06,Hamaguchi06}. Specifically, it
should be possible to measure the stau mass with an accuracy of $\sim 1$~\%.
However, most of the LHC staus would be produced with $\beta \gamma \sim 1$, and
would decay outside the detector if they have lifetimes $> 10^{-7}$~s.
We recall that there are no significant cosmological constraints on such staus
if their lifetimes $< 1$~s, and that there are only weak constraints on
staus with lifetimes $< 10^3$~s. Therefore, collider information on the stau
lifetime would be interesting for cosmology as well as the theory of
supersymmetry breaking. However, the numbers of slow staus that
stop inside the LHC detectors would be relatively modest: perhaps just a handful
in benchmarks $\eta$ and $\zeta$, and somewhat more in benchmark $\epsilon$.
The LHC would detect no staus in the case of benchmark $\theta$.

If the stau is indeed found by the LHC, in these scenarios one would know the optimal
energy for producing the optimal number of stopping staus via the reaction
$e^+ e^- \to \tilde{\tau}_{1}^+ \tilde{\tau}_{1}^-$. As we have
emphasized in this paper, there would in general be many
additional staus coming from the decays of heavier sparticles produced by reactions
such as $e^+ e^- \to \tilde{e}_R \tilde{e}_L$, $e^+ e^- \to \chi \chi$, etc.. The masses
of these heavier sparticles could also be measured at the LHC in models resembling
benchmarks $\epsilon, \eta$ and $\zeta$, and hence the corresponding $e^+ e^-$
thresholds could be estimated. These production mechanisms would also yield
slow staus that should be included in the choice of the $e^+ e^-$ center-of-mass
energy in order to maximize the number of stopped staus.

Alternatively, the stau might be too heavy to be produced at the LHC, as in
benchmark scenario $\theta$ which, we recall, has the added merit of being
$^{6,7}$Li-friendly. In such a case, one would need to measure the stau mass in
$e^+ e^-$ annihilation itself before going to the optimal energy for stopping staus.
This would be depend on the masses of heavier sparticles, which could be
determined either directly or by invoking some model.

The signature for a stopped stau would be its decay
$\tilde{\tau}_1\rightarrow\tilde{G}\tau$ out of coincidence with a
collision in the central detector~\cite{Martyn06}. The $\tau$ decay modes (and their
branching ratios) observable in the collider detector would be the
leptonic 3-body decays $\tau\rightarrow \nu_{\tau}\mu\nu_{\mu}$
(17.4\%) and $\tau\rightarrow \nu_{\tau}e\nu_{e}$ (17.8\%), and
hadronic decays $\tau\rightarrow \nu_{\tau}\pi$ (11.1\%),
$\tau\rightarrow\nu_{\tau}\rho\rightarrow\nu_{\tau}\pi^{\pm}\pi^{0}$
(25.4\%) and
$\tau\rightarrow\nu_{\tau}3\pi\rightarrow\nu_{\tau}\pi^{\pm}\pi^{+}\pi^{-}
+\nu_{\tau}\pi^{\pm}\pi^{0}\pi^{0}$ (19.4\%). In the benchmark
scenarios considered here, the decaying $\tau$s would be very
energetic, since $m_{\tilde G} (= 20, 100, 20, 66~{\rm GeV}) <<
m_{\tilde \tau_1} ( = 154, 346, 327, 1140~{\rm GeV})$,
respectively. These $\tau$ decays are therefore likely to produce
isolated, high-energy hadronic or
electromagnetic clusters above the threshold of the HCAL ($E_{h,em}>10$ GeV),
hadronic showers in the yoke ($E_{h}>10$ GeV), or energetic
$\mu$s originating in the HCAL or yoke ($E_{\mu}>10$ GeV)~\cite{Martyn06}. The main
background would come from cosmic rays, and some rejection could be obtained by
excluding decay vertices in the outermost detector layers and vetoing
signals initiated by external muons. As pointed out in~\cite{Martyn06}, further excellent
background discrimination would be provided by requiring the decay
vertex to be consistent with the estimated stopping point of a
$\tilde{\tau_{1}}$ detected and measured previously. The $\tau$
recoil energy
$E_{\tau}=m_{\tilde{\tau}}\left(1-m_{\tilde{G}}^{2}/m_{\tilde{\tau}}^{2}\right)/2$
would have sensitivity to the gravitino mass~\cite{Martyn06}. We note
in addition that the tau decay
spectra measurable in each of the above decay channels would be
sensitive to the tau polarization and hence the handedness of the
parent stau.

This analysis demonstrates once again the complementarity of the LHC and a linear
electron-positron collider of sufficiently high energy. Information from the LHC may not
only establish the threshold for new physics, but also provide an estimate of the optimal energy
for studying it. At least for certain studies, such as the decays of stopped staus considered here,
this optimal energy may be significantly higher than the $e^+ e^-$ production threshold.

\section*{Acknowledgements}

We would like to thank Albert De Roeck, Filip Moortgart and Daniel
Schulte for valuable discussions and comments on the subject. The
work of O.C. and Z.K. was supported in part by the Turkish
Atomic Energy Authority (TAEK) under the grants no VII-B.04.DPT.1.05
and Turkish State Planning Organization
(DPT) under the grants no DPT-2006K-120470.

\end{document}